\documentclass[fleqn,usenatbib]{mnras}

\usepackage{newtxtext,newtxmath}
\usepackage[T1]{fontenc}
\usepackage{ae,aecompl}
\usepackage{times}

\usepackage{rotating,graphicx}	
\usepackage{amsmath}	
\usepackage{amssymb}	
\usepackage{ulem}
\usepackage[ampersand]{easylist}

\usepackage{latexsym}
\usepackage{enumerate}
\usepackage{amstext}
\usepackage{morefloats}

\usepackage{natbib}
\usepackage{rotating}
\usepackage{changepage}
\usepackage{scrextend}
\usepackage[ampersand]{easylist}
\usepackage{xcolor}
\usepackage{ulem}

\usepackage{mathtools}

\title[Star Cluster Stellar Wind Retention \& Expulsion]{Should I Stay or Should I Go:  Stellar Wind Retention and Expulsion in Massive Star Clusters}

\author[J.~P.~Naiman et al.]{J.~P.~Naiman$^{1}$\thanks{E-mail: jill.naiman@cfa.harvard.edu}, E.~Ramirez-Ruiz$^{2}$, D.~N.~C.~Lin$^{2}$ 
\vspace*{0.2cm} \\
$^{1}$Harvard-Smithsonian Center for Astrophysics, 60 Garden Street, Cambridge, MA, 02138, USA\\
  $^2$Department of Astronomy and Astrophysics, University of California, Santa Cruz, CA 95064, USA
}

\date{11 July 2017}

\pubyear{2017}

\begin{document}
\label{firstpage}
\pagerange{\pageref{firstpage}--\pageref{lastpage}}
\maketitle

\begin{abstract}
Mass and energy injection throughout the lifetime of a star cluster contributes to the gas reservoir available for subsequent episodes of star formation and the feedback energy budget responsible for ejecting material from the cluster.
In addition, mass processed in stellar interiors and ejected as winds has the potential to augment the abundance ratios of currently forming stars, or stars which form at a later time from a retained gas reservoir.
Here we present hydrodynamical simulations that explore a wide range of cluster masses, compactnesses, metallicities and stellar population age combinations in order to determine the range of parameter space conducive to stellar wind retention or wind powered gas expulsion in star clusters.
We discuss the effects of the stellar wind prescription on retention and expulsion effectiveness, using MESA stellar evolutionary models as a test bed for exploring how the amounts of wind retention/expulsion depend upon the amount of mixing between the winds from stars of different masses and ages.
We conclude by summarizing some implications for gas retention and expulsion in a variety of compact ($\sigma_v \gtrsim 20 \, {\rm km s^{-1}}$) star clusters including young massive star clusters ($10^5 \lesssim M/M_\odot \lesssim 10^7$, $age \lesssim 500$~Myrs), intermediate age clusters ($10^5 \lesssim M/M_\odot \lesssim 10^7$, $age \approx 1-4$~Gyrs), and globular clusters ($10^5 \lesssim M/M_\odot \lesssim 10^7$, $age \gtrsim 10$~Gyrs).
\end{abstract}



\section{Introduction}

 Whether and when gas is retained within a star cluster, along with inter-cluster and intra-cluster gas heating mechanisms determines the star formation history of the cluster.
 Many cluster properties can be responsible for gas retention or expulsion in a cluster - the cluster's compactness, cluster mass, ability to cool effectively, mass and temperature of gas reservoir, and efficiency of internal and external heating mechanisms.  
 As it thought that most stars form in clusters \citep[e.g.][]{lada2003}, the study of which parameters and mechanisms are important at different cluster evolutionary times is crucial to understand when favorable conditions for star formation arise throughout a cluster's lifetime.

 Observations of young massive clusters (YMCs), intermediate age clusters (IACs) and globular clusters (GCs) potentially give us insights into gas retention at different times within massive star clusters \citep[e.g.][]{bastian2017}.
 Additionally, if GCs evolve from YMCs and IACs as some have suggested \citep{longmore2014,krui2015} then observations of each type of system provides clues to the origin of the high level of occurrence of multiple subpopulations with different abundance ratios within present day GCs.

 Variations in the abundances of He, Mg, C, N, O, Mg, Na and Al in main sequence stars \citep{gratton2001,briley2002,cohen2002,cannon1998,pancino2010b} and RGB stars \citep{sneden2004} within the majority of Galactic globular clusters suggest a complex enrichment history during the formation of stars within these systems.
 In addition, the observed anti-correlations between several elements \citep[Na-O and Mg-Al;][]{kraft1993,ivans1999,carretta2006b,carretta2009b,conroy2012} require a significant fraction of the enriching material to be processed at temperatures $> 10^7 \, {\rm K}$ in order for the CNO, Na-Ne and Mg-Al cycles to be activated \citep{karakas2007,ventura2008b}.
 As the anomalous population is responsible for 30-70\% of the stellar content of the globular clusters \citep[e.g.][]{carretta2009} any scenario invoked to explain this complex star formation history must also account for the pollution of a significant portion of the stellar population or the removal of the majority of the unpolluted first generation of stars.
 Additionally, the spatial distributions of anomalous stars, individual stellar abundance spreads, limits on the initial mass of globular clusters and various other constraints must be considered in any process capable of producing multiple episodes of star formation in YMCs if they are indeed the progenitors of current globular clusters \citep{bastian2013b,krui2015}.

 Several mechanisms have been proposed to explain these abundance variations.
 While some rely directly on gas retention and protracted enrichment of populations of stars formed after the initial burst of star formation  \citep{prantzos2007,ventura2008a,ventura2008b,pflamm2009,dercole2010,conroy2011a} others rely on enrichment processes which occur during one main star formation event \citep{bastian2013b,deMink2009,den2014,prantzos2006}.
Whether directly or indirectly, stellar winds can play an important role in the retention or expulsion of gas during enrichment processes - if hot stellar winds thermalize efficiently with other sources of gas they can drive material out of a star cluster, while if cold stellar winds are added to the intercluster medium they can aid in the rapid cooling of the gas and help to trigger star formation.
 Thus, a careful study of gas retention and expulsion from stellar winds is a necessary component of understanding what cluster potential parameters and ages are favorable for gas retention, and what combinations lead to an overall expulsion of gas from the system.

 In the context of wind retention in proto-globular clusters, there have been several one dimensional simulations of gas over a small range of wind and cluster parameters, particularly for slow AGB winds  ($v_w = 10 \, km s^{-1}$), in the presence of supernovae and star formation \citep[e.g.][]{dercole2008}.
 Additionally, a parametric study of  the role  of cluster mass and compactness has been done, albeit using only  a  limited range of structural parameters \citep{vesperini2010}.
 There have also been several detailed three dimensional studies focused on stellar wind retention/expulsion in star clusters (GCs; \citealt{calura2015,priestley2011}, T Tauri/Herbig Ae/Be star clusters; \citealt{rg2008}, OB clusters; \citealt{pittard2013}), however, to study the interaction of a wide range of stellar wind parameters from various stellar population ages along with a variety of stellar cluster masses and compactnesses the required full simulation suite over all parameter space in three dimensions would be computationally prohibitive.

 Because of their shallow potentials, careful treatment of stellar winds within star clusters is essential to properly model their gas retention and expulsion abilities in both one and three dimensions.
 Many models assume and rely on slow winds from the interiors of AGB envelopes to drive large amounts of stellar wind retention in the cores of young massive clusters \citep[e.g.][]{dercole2008}.
 However, observations of fast winds from the upper envelope layers of evolved stars complicate this picture \citep{mauas2006}.
 In addition, thermalization of evolved stellar winds with the faster winds of the more numerous main sequence stars may be effective at expelling gas from stellar clusters \citep{smith1999}.

 Adding to the complexity of the evolution of these systems is the existence of many other mechanisms beyond stellar winds that can be responsible for the deposition of mass into massive star clusters at different times in their evolutionary history.
 If the cluster does not form in isolation, accretion of external gas could pollute enriched material and thus change the abundance patterns of forming stars \citep{pflamm2009,dercole2010,naiman2011,conroy2012}.
 If this non-isolated cluster moves too quickly through its background medium, ram pressure stripping could remove the majority of the gas from these systems, effectively quenching star formation \citep{naiman2009,priestley2011}.
 Additionally, any low level gas produced later in the star cluster's life may be stripped from passages through its host galaxy's disk, depending on its particular orbital parameters \citep{desilva2009,martell2009,pancino2010a}.

 Early in a star cluster's evolution there are a multitude of additional sources of energy injection that might be responsible for heating the intercluster gas.
 Both SNII and SNIa can add energy and heavy elements to a star cluster in the first tens to hundreds of Myrs of a star's lifetime, and, under certain conditions, can effectively drive all gas out of these young clusters \citep{krause2013}, while classical novae can drive out gas in all but the most massive clusters at slightly later times \citep{moore2011}.
 Accretion onto compact stellar remnants is another possible avenue to clear gas rapidly in young star clusters \citep{leigh2013a}.
 Lyman-Werner flux serves as a means to heat intercluster gas and keep stars from forming during the first few Myrs of a star cluster's life \citep{conroy2011b,conroy2012,krause2013}, and gas heating by UV radiation from a population of white dwarfs serves as a possible avenue to expel intercluster gas at late times \citep{mcdonald2015}.
 Indeed, there exists many means to expel gas from recently formed clusters to distances on the order of hundreds of parsecs \citep{bastian2014b,hollyhead2015}.  

 In the work presented here we focus on predominately using stellar winds as the sole source of mass and energy in the formation of isolated star clusters and only briefly touch on the effects of extra sources of mass and energy.
 Throughout the majority of this paper we remain agnostic about the origin of the abundance spreads observed in present day GCs and their possible evolution from YMCs and IACs, and instead endeavor to study the role that the mass loss and energy injection from stellar winds plays in determining whether significant gas is retained or expelled from a star cluster given the cluster's mass, compactness, and age.
 The paper is outlined as follows:
 In section \ref{section:hydro} we review our implemented hydrodynamical scheme.  Section \ref{section:stellarEvolution} describes our stellar evolutionary models generated with the MESA \citep{paxton2011} code, paying careful attention to the prescriptions used to describe the mass loss and wind velocities emanating from stars of different masses.  This section also describes how the results for individual stars of different masses are combined into average mass and energy loss prescriptions from a population of stars, under different assumptions about the effective thermalization of winds from different stellar populations.  In section \ref{section:results} we present the resultant stellar wind mass retention for star clusters with a variety of masses, compactnesses and ages.  Section \ref{section:discussion} briefly discusses the effects of other heating mechanisms beyond those from stellar winds on the removal of material from star clusters and touches on the effects of different gas metallicities on the cooling properties of intercluster gas.  Finally, we end with a discussion of the application of our results to gas retention and star formation in present day YMCs, IACs, and GCs presented in section \ref{section:applications} and summarize our conclusions in section \ref{section:conclusions}.

\section{Hydrodynamical Methods and Initial Setup} \label{section:hydro}

 To examine the ability of clusters to effectively retain the winds emanating from their evolving stellar members, we simulate mass and 
energy injection in isolated core potentials under the assumption of spherical symmetry \citep{quad2004,hue2010}.
 The one-dimensional  hydrodynamical equations are solved using 
FLASH, a parallel, adaptive mesh
refinement hydrodynamical code \citep{fryxell2000}. The winds from the closely packed stellar members are assumed to shock and 
thermalize such that  density and energy contributions can be treated as source terms in the hydrodynamical 
equations.  

 The stellar cluster potentials are modeled as Plummer models \citep{bruns2009,pflamm2009}
\begin{equation}
\Phi_g = - \frac{G M_c}{\left[r^2 + r_c^2(\sigma_v) \right]^{1/2}}
\end{equation}
for a  given total mass, $M_c$, and velocity dispersion, 
$\sigma_v = \left(3^{3/4}/ \sqrt{2}\right)^{-1} \sqrt{G M_c/r_c}$.  
While the shape of the potential can have an effect on the radial distribution of gas  inside the core 
\citep{pflamm2009,naiman2011}, the overall amount of gas accumulated in the cluster  is relatively unchanged 
by the exact form of its potential.  For the sake of simplicity, we assume the potential does not change appreciably over the period of gas accumulation.

In addition to following the dynamics of the 
gas under the influence of $\Phi_g$, 
the self gravity of the gas is computed using FLASH's multipole gravity module.  
We fix our resolution to 6400 radial cells for each model, with linearly equidistant spacing between the minimum ($r_{\rm min}$) and maximum ($r_{\rm max}$) radius. 
The resolution within the computational domain  is 
set by the core radius, $r_{\rm c}$, to ensure that we are adequately resolving the 
core  and that the cluster's potential is effectively zero at the outer boundary (cell centers are then approximately $r_{\rm min} \sim 0.02 \, r_{\rm c}$, and $r_{\rm max} \sim 65 \, r_{\rm c}$).

 Given this potential and assuming spherical symmetry, at a time $t_{n}$, we construct the conservation of mass, momentum, and energy in a more general from starting from those in \citep{hue2010,priestley2011}:
\begin{equation}
\begin{multlined}
\rho(r,t_n) = \rho(r,t_{n-1}) + q_{m}(r,t_{n}) dt(t_n,t_{n-1})
\end{multlined}
\label{eq:massconsLong}
\end{equation}

\begin{equation}
\begin{multlined}
v(r,t_{n}) =  v(r,t_{n-1}) \rho(r,t_{n-1})/\rho(r,t_{n}) + a_g(r,t_{n+1})dt(t_n,t_{n-1}) \\
{+ p_{\rm non-uniform}(r_{i-1} ,r_{i+1})/\rho(r,t_{n}) + p_\sigma(r_{i+1},r_{i-1})/\rho(r,t_{n})}
\end{multlined}
\label{eq:momentumLong}
\end{equation}

\begin{equation}
\begin{multlined}
\rho(r,t_n)\epsilon(r,t_n) = \frac{1}{2}\rho(r,t_{n-1})v(r,t_{n-1})^2 + \rho(r,t_{n-1})\epsilon(r,t_{n-1})  \\
- \frac{1}{2}\rho(r,t_{n})v(r,t_{n})^2 + q_\varepsilon(r,t_{n-1})dt(t_n,t_{n-1})-Q(r,t_{n-1})
\label{eq:energyLong}
\end{multlined}
\end{equation}
 where $ \rho(r,t)$, $u(r,t)$ and $\varepsilon(r,t)$ are the gas density, pressure,  radial velocity and internal energy density, respectively.    
The term $a_g(r,t)$ includes the gravitational effects of both the stellar cluster potential ($\Phi_g$) and the self gravity of the gas.
 Here,  $Q(r,t) = n_i(r,t) n_e(r,t) \Lambda(T,Z)$ is the cooling rate for gas with ion and electron number densities 
$n_i(r,t)$ and $n_e(r,t)$, and $\Lambda(T,Z)$ is the cooling function for gas at temperature $T$ and 
metallicity $Z$.  We use $\Lambda(T,Z)$ from \cite{gnat2007} for $T > 10^4 \, {\rm K}$ and 
\cite{dalgarno1972} for $10 \le T \le 10^4 \, {\rm K}$.
 
 The $q_m(r,t)$ and $q_\varepsilon(r,t)$ terms in equations (\ref{eq:masscons})-(\ref{eq:energy}) are used here to mediate  the total rate of mass and energy injection at a time $t$ in a cluster's history.
For $N$ stars each with average mass loss rate {\bf at time $t$ of} $\langle \dot{M}(t)\rangle$ and wind energy injection rate $\frac{1}{2} \langle \dot{M}(t)\rangle \langle v_{\rm w}(t)^2\rangle$ 
we have 
\begin{equation}
\dot{M}(t)_{\rm w,total} =  N \langle \dot{M}(t)\rangle = \int 4 \pi r^2 q_m(r,t) dr 
\label{eq:mdotgen}
\end{equation}
 and 
\begin{equation}
\dot{E}(t)_{\rm w, total} = \frac{1}{2} N \langle \dot{M}(t) \rangle \langle v_{\rm w}(t)^2\rangle = \int 4 \pi r^2 q_\varepsilon(r,t) dr
\label{eq:kegen}
\end{equation} 
such that $q_\varepsilon(r,t) = \frac{1}{2} q_m(r,t) \langle v_{\rm w}(t)^2 \rangle$.   
 For simplicity, we neglect the effects of mass segregation on the positions of main sequence and evolved stars and assume 
$q_m(r,t) \propto n(r)$ such that $q_m(r,t) = A(t) r^{-2} \frac{d}{dr} \left(r^2 \frac{d\Phi_g}{dr}\right)$.  Here, $A(t) = \langle \dot{M}(t) \rangle / (4 \pi G \langle M_\star \rangle)$ with the average mass of a star given by $\langle M_\star \rangle$.

 The momentum term accounting for a non-uniform distribution of stellar wind ejected momentum within the cluster, $p_{\rm non-uniform}$ can be derived assuming the momentum flowing into (out of) the $i$th cell is the integral over the momentum from the sides of stars facing outward (inward) in the $i-1$ ($i+1$) cell: $d\vec{p}_{{\rm non-uniform},\pm} = \pm q_m(r_{i-1/i+1},t) v_w dt/d\Omega \hat{r}$, where $\hat{r}$ is the unit vector pointing radially outward from the cluster's center.  This assumes the momentum flux carried by the wind from each individual star is isotropic. Integrating over the outward (inward) solid angles assuming the spherical symmetry of the individual stellar winds gives $\vec{p}_{\rm{non-uniform},\pm} = \pm \frac{1}{2} q_m(r_{i-1/i+1},t) v_w dt \hat{r}$.  Given our expression for $q_m$, for the $i$th cell this is expressed as:
\begin{equation}
p_{\rm non-uniform}(r_i) = 
v_w f_{\rm c} \left[G(r_{i-1},r_{\rm c}) - G(r_{i+1}, r_{\rm c})  \right] dt
\label{eq:nonuni}
\end{equation}
where
\begin{equation}
G(r_j, r_{\rm c}) = \left( 1 + \frac{r_{j}^2}{r_{\rm c}^2} \right)^{-5/2} 
\end{equation}
and
\begin{equation}
f_{\rm c} = \left( \frac{3 M_{\rm c}}{8 \pi r_{\rm c}^3 \langle M_\star \rangle} \right)
\end{equation}

 The second of the extra momentum terms, $p_\sigma$, arrises from the fact that the stars are moving, and because the distribution of stars is not uniform, this extra momentum is not distributed uniformly throughout the cluster.  Following a procedure similar to deriving equation \ref{eq:nonuni} with $d\vec{p}_{\sigma,\pm} = \pm q_m(r_{i-1/i+1},t) \sigma(r_{i-1/i+1}) dt/d\Omega \hat{r}$ and $\sigma = \sqrt{G M(<r)/r}$ this momentum term can be expressed as:
\begin{equation}
p_\sigma(r_i) = \sqrt{\frac{G M_{\rm c}}{r_{\rm c}^3}} f_{\rm c}
\left[F(r_{i-1}, r_{\rm c}) - F(r_{i+1}, r_{\rm c})\right] dt
\end{equation}
where
\begin{equation}
F(r_j, r_{\rm c}) =  r_j \left( 1 + \frac{r_j^2}{r_{\rm c}^2} \right)^{-13/4}
\end{equation}

 Instead of $p_{\rm non-uniform}$ and $p_\sigma$, many authors include an extra term in the energy equation \ref{eq:energyLong} on the order of $\frac{1}{2} n(r) \sigma^2$, where $\sigma$ is the velocity dispersion of the stars within the cluster and $n(r)$ is the stellar density to account for the fact that the stars are moving \citep{faulk1977,smith1999,dercole2008,priestley2011}. 
 In compact clusters, this extra energy injection term heats the gas, potentially delaying star formation until enough mass has been accumulated in the central regions to cool efficiently and subsequently trigger star formation.
 However, including such a term without accounting for how this energy loss would slow a star's motion violates conservation of energy over the lifetime of the star cluster and therefore we do not include it here, instead relying on $p_\sigma$ to include the effects of stellar motion in our prescription. 

 A final source of energy from stars orbiting within our clusters arrises from the interaction of stellar motion induced shocks and stellar winds.
 The energy injection rate for shocked gas of density $\rho_{\rm g}$ around a star of mass $M_\star$ moving at velocity $\sigma_v$ is given by $\dot{E}_{\rm s} \approx (1/2) \rho_{\rm g} \sigma_v^2 \Re$.  
 Here, the gas interaction rate at the bow shock generated by the moving star, $\Re = \sigma_v \Sigma$, can be estimated from the bow shock radius of the star \citep{wilkin1996}, $\Sigma \approx \pi R_0^2 = \pi \dot{M} v_w/(4 \rho_g \sigma_v^2)$ where $\dot{M}$ is the mass loss from a single star at stellar wind velocity $v_w$.
 Assuming the energy injection from a star's stellar wind is given by  $\dot{E}_{\rm w} = (1/2) \dot{M} v_w^2$, then the ratio of injection rates can be written simply as 
\begin{equation}
\dot{E}_{\rm s}/\dot{E}_{\rm w} \approx \sigma_v/(4 v_w).
\label{eq:esew}
\end{equation}
 Thus, appreciable changes in the heating rates caused by the motion of the stars are only expected when the cluster velocity dispersion is  larger than $4v_w$. 
 For the cluster velocity dispersions discussed here, this term is negligible and thus it is not included.

 Additionally, we find the $p_\sigma$ and $p_{\rm non-uniform}$ terms contribute to the momentum equation minimally. As both terms are proportional to $q_m$ and either $v_w$ or $\sigma_v$, it is expected that they should contribute to the hydrodynamics only at early times with the intercluster density, $\rho_{\rm gas}$ is much smaller than $q_m dt$, in clusters which can only retain very small amounts of material ($\rho_{\rm gas} << q_m dt$ for all time), and for very fast winds and compact clusters.  For the parameters we will be concerned with in this paper we find these effects are of the order of a few percent. While both terms aid, albeit minimally, to removing material from the cluster at large radii, within the core the $p_\sigma$ term contributes to funneling gas into the center of the star cluster.

 Once we drop the minimally contributing momentum terms from equations (\ref{eq:momentumLong}) and (\ref{eq:energyLong}), we revert to the equations typically seen in other works \citep{priestley2011,hue2010}:
\begin{equation}
\rho(r,t_n) = \rho(r,t_{n-1}) + q_{m}(r,t_{n}) dt(t_n,t_{n-1})
\label{eq:masscons}
\end{equation}
\begin{equation}
v(r,t_{n}) = v(r,t_{n-1}) \rho(r,t_{n-1})/\rho(r,t_{n}) + a_g(r,t_{n+1})dt(t_n,t_{n-1})
\label{eq:momentum}
\end{equation}
\begin{equation}
\begin{multlined}
\rho(r,t_n)\epsilon(r,t_n) = \frac{1}{2}\rho(r,t_{n-1})v(r,t_{n-1})^2 + \rho(r,t_{n-1})\epsilon(r,t_{n-1}) \\
- \frac{1}{2}\rho(r,t_{n})v(r,t_{n})+ q_\varepsilon(r,t_{n-1})dt(t_n,t_{n-1})-Q(r,t_{n-1})
\label{eq:energy}
\end{multlined}
\end{equation}

 We also neglect the effects of external mass accumulation, which can be an important source of gas when 
clusters reside in cool, dense ISM gas \citep{naiman2009,naiman2011,priestley2011,conroy2012}.
In addition to ignorning shock heating of the gas caused by  the motion of the stars derived in equation \ref{eq:esew} we ignore other specific heating sources such as photoionization or  supernova, but discuss the overall effects of additional energy injection in the cluster in  \S\ref{section:discussion}.  

 In cases where catastrophic cooling occurs, we assume star formation is triggered if the Jeans length of the collapsing gas is smaller than the 
central resolution element, or if $t_{\rm cool} \lesssim t_{\rm dyn}$.  To estimate the gas evolution during the star forming period, our cooling prescription is modified following \citet{truelove1997}  by turning off  energy losses  and allowing the gas to evolve adiabatically.  
Such a method approximates the transition of an isothermally collapsing cloud to an optically thick, adiabatically 
evolving protostellar cluster while forgoing the computationally expensive three dimensional radiative transfer calculations 
required to treat this problem accurately.
 Because we do not include an explicit star formation prescription, we allow such unstable regions to evolve  for  a few sound crossing times 
before halting the simulation.  With this implementation, a decrease in density of approximately one order of magnitude or more occurs without the influence of cooling.  While this decrease results in a significant suppression of mass accumulation within the cluster, some gas may be retained until it is either stripped by external mechanisms or displaced by supernovae from the second generation of stars, approximately 10~Myrs after they form.\\
\\

\section{Stellar Evolution and Mass Loss on the HR Diagram} \label{section:stellarEvolution}

The final ingredients to be specified in our simulations are the time dependent average mass loss rates and wind velocities, which in turn determine the mass and energy injection rates, $q_m$ and $q_\varepsilon$.  Because our simulations are spherically symmetric, these rates must encompass the average mass loss properties of the stellar population as a whole.

\subsection{Mass Loss Rates and Wind Velocities from Individual Stars: Prescriptions vs. Observations}

 To compute the individual mass loss rates, $\dot{M}(M_\star,t)$,  and wind velocities, $v_w(M_\star,t)$,  as both a function of the mass of the star, $M_\star$ and its age, $t$, we use the MESA stellar evolution models \citep{paxton2011}.
MESA is used to follow the evolution of a grid of stellar models with $M_\star = 0.1 \, M_\odot$ - 
$20 \, M_\odot$ from ZAMS to either the white dwarf stage, end of the AGB phase, or 
compact object creation.

 The mass and kinetic energy injection can vary by orders of magnitude throughout a star's lifetime, therefore, to constrain our MESA models we compare our calculated mass loss rates and wind velocities with observed quantities throughout the HR diagram.

\subsubsection{Mass Loss Rates} \label{section:masslossrates}

The majority of mass loss from stellar winds occurs in the RGB and AGB phases of low and intermediate mass stars ($1 \lesssim M/M_\odot \lesssim 8$) making this timeframe of mass loss important to constrain.
In MESA the mass loss rates during the crucial AGB phase are implemented using 
the wind prescription of  \citet{blocker1995}, $\dot{M}_B = 1.93 \times 10^{-21}  \eta_B (M/M_\odot)^{-3.1} (L/L_\odot)^{3.7} (R/R_\odot) \, M_\odot {\rm yr^{-1}}$, with a normalization of $\eta_B = 0.04$ consistent with LMC measurements \citep{ventura2000} 
and with previous studies \citep{dercole2008,dercole2010,conroy2011b}.  
 The \cite{reimers1975} prescription, $\dot{M}_R = 4 \times 10^{-13}  \eta_R (M/M_\odot)^{-1} (L/L_\odot) (R/R_\odot) \, M_\odot {\rm yr^{-1}}$, $\eta_R = 1.0$,  is used in MESA to estimate the mass loss during the RGB branch.   

\begin{figure}
\centering\includegraphics[width=0.35\textwidth, angle=90]{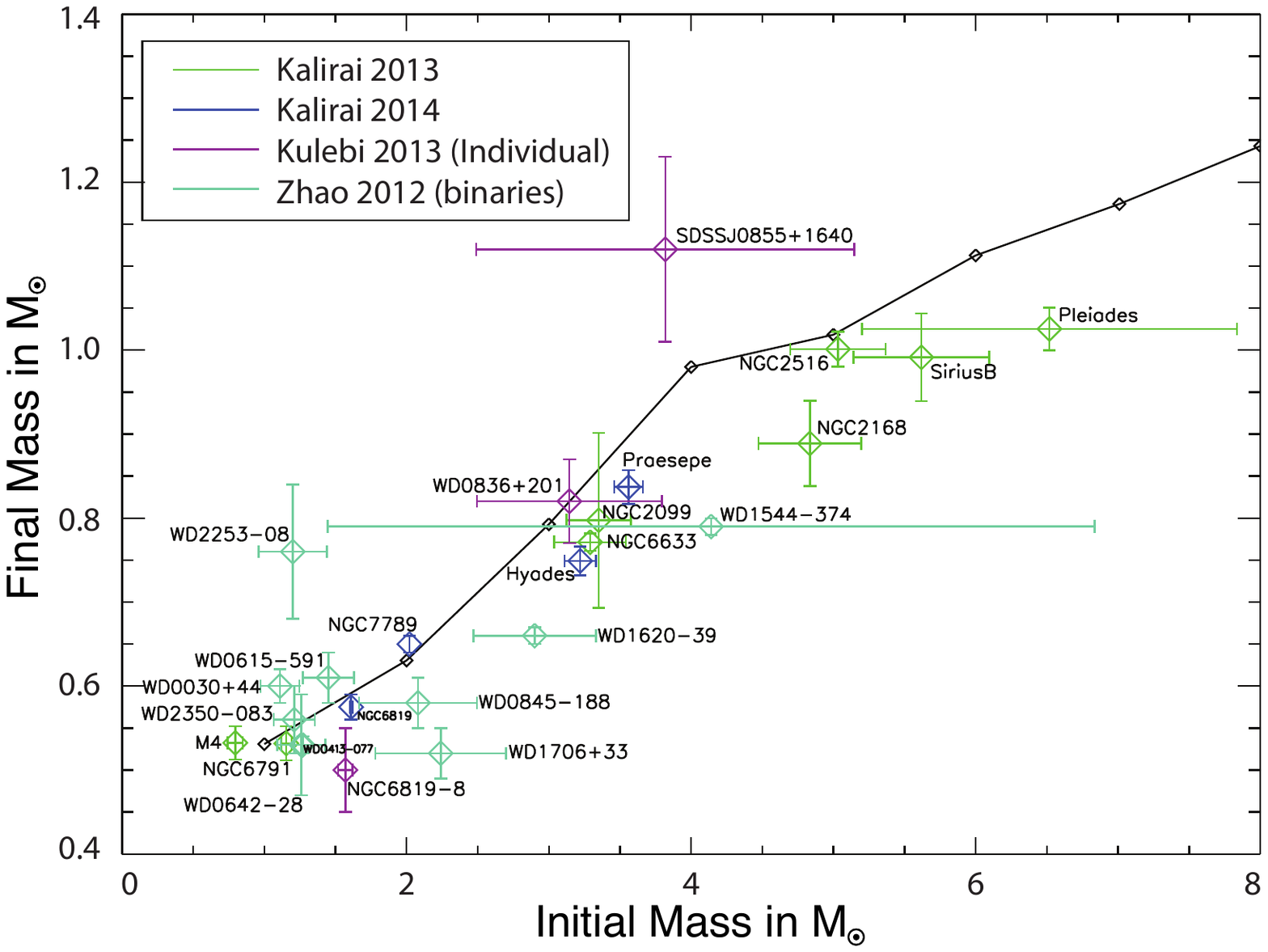}
\caption{ Initial final mass relation from observations and MESA calculations. The \citealt{kalirai2013a} and \citealt{kalirai2013b} points are averaged over the total observed white dwarfs in each respective cluster, while the \citealt{kulebi2013} and \citealt{zhao2012} points are taken from individual observations.}
\label{fig:ifrelation}
\end{figure}

In general, the mechanisms that drive mass loss during the RGB and AGB phases are not well constrained \citep{ventura2008b,karakas2007,marigo2012}, and as such, it is generally unproductive to compare instantaneous mass loss rates between our models and observations.
However, we can constrain mass loss prescriptions during these crucial phases by comparing the total mass lost by the MESA models with the observed the initial-final mass relation.   As shown in Figure \ref{fig:ifrelation}, we find that our models reproduce the initial-final mass relation over this key mass range.

\begin{figure}
\centering\includegraphics[width=0.46\textwidth]{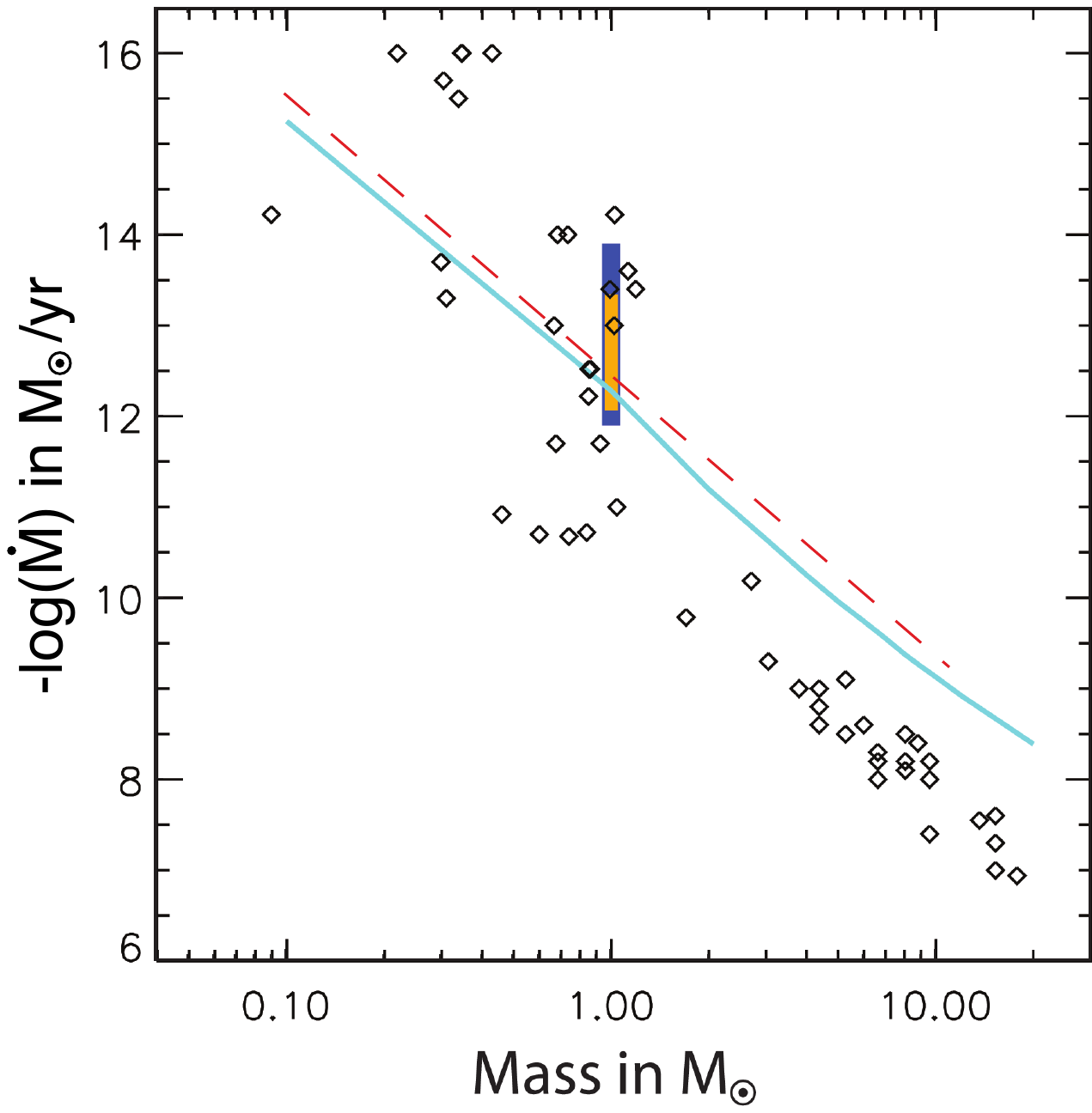}
\caption{Mass loss rate estimates along the main sequence as a function of initial stellar mass.  Observational estimates are plotted as black diamonds \citep{cranmer2011,dejager1988,searle2008,waters1987,debes2006,badalyan1992,morin2008}.  The blue rectangle denotes the estimated variations in the solar mass loss rate \citep{wood2005}, and the yellow rectangle shows the observed variations in the mass loss rate of the Sun as a function of its X-ray activity \citep{cohen2011}.  The blue line shows the main sequence mass loss rate prescription used by MESA and the often derived analytical prescription is shown with the dashed red line. }
\label{fig:msmassloss}
\end{figure}

While mass loss along the main sequence is minimal, the kinetic energy injection can be significant as will be discussed in detail in section \ref{section:averages}.
Therefore, to account for mass loss during this stellar evolutionary phase, we extend the \cite{reimers1975} prescription to mass loss along the main sequence in our MESA models - $\dot{M}_{R,MS} = 4 \times 10^{-13}  \eta_R (M/M_\odot)^{-1} (L/L_\odot) (R/R_\odot) \, M_\odot {\rm yr^{-1}}$, with $\eta_R = 1.0$.  
 This prescription fits within the highly variable range of mass loss rates observed in main sequence stars as depicted in Figure \ref{fig:msmassloss}.  

We end this section by noting that the main sequence mass loss rate may also be estimated analytically using the typical relations between mass, luminosity and radius along the main sequence.  Using the Reimers mass loss prescription and the main sequence relations of \cite{demircan1991} for luminosity, $L \simeq 1.03 (M/M_\odot)^{3.42} \, L_\odot$, and radius $R \simeq 0.85 (M/M_\odot)^{0.67} \, R_\odot$,
the analytical relation for mass loss along the main sequence, $\dot{M}_{R,MS} \simeq \eta_R 3.5 \times 10^{-13}(M/M_\odot)^{3.09} \, \, M_\odot/yr$, is plotted in Figure \ref{fig:msmassloss}.  In general, the MESA relation and analytical fits are comparable, but we caution this will change based on the various iterations of the equations from \cite{demircan1991}.

\subsubsection{Wind Velocities}

To include the full effects of the material ejected by unbound stellar winds in our simulations knowledge of not only the mass loss rates of the stars, but the amount of kinetic energy injected by winds during different phases of evolution for each star is necessary.  Here, we take the wind velocity, $v_w$, coupled with the mass loss rate, as a proxy for the amount of energy injected by stellar winds - $E_{\rm th,winds} \sim \dot{M} v_{\rm w}^2$.
This assumption requires efficient thermalization between the stellar wind ejecta and intercluster gas.  Thus, in order to proceed it is necessary to determine the velocity of stellar wind material which is unbound to the mass losing stars.

\begin{figure}
\centering\includegraphics[width=0.46\textwidth]{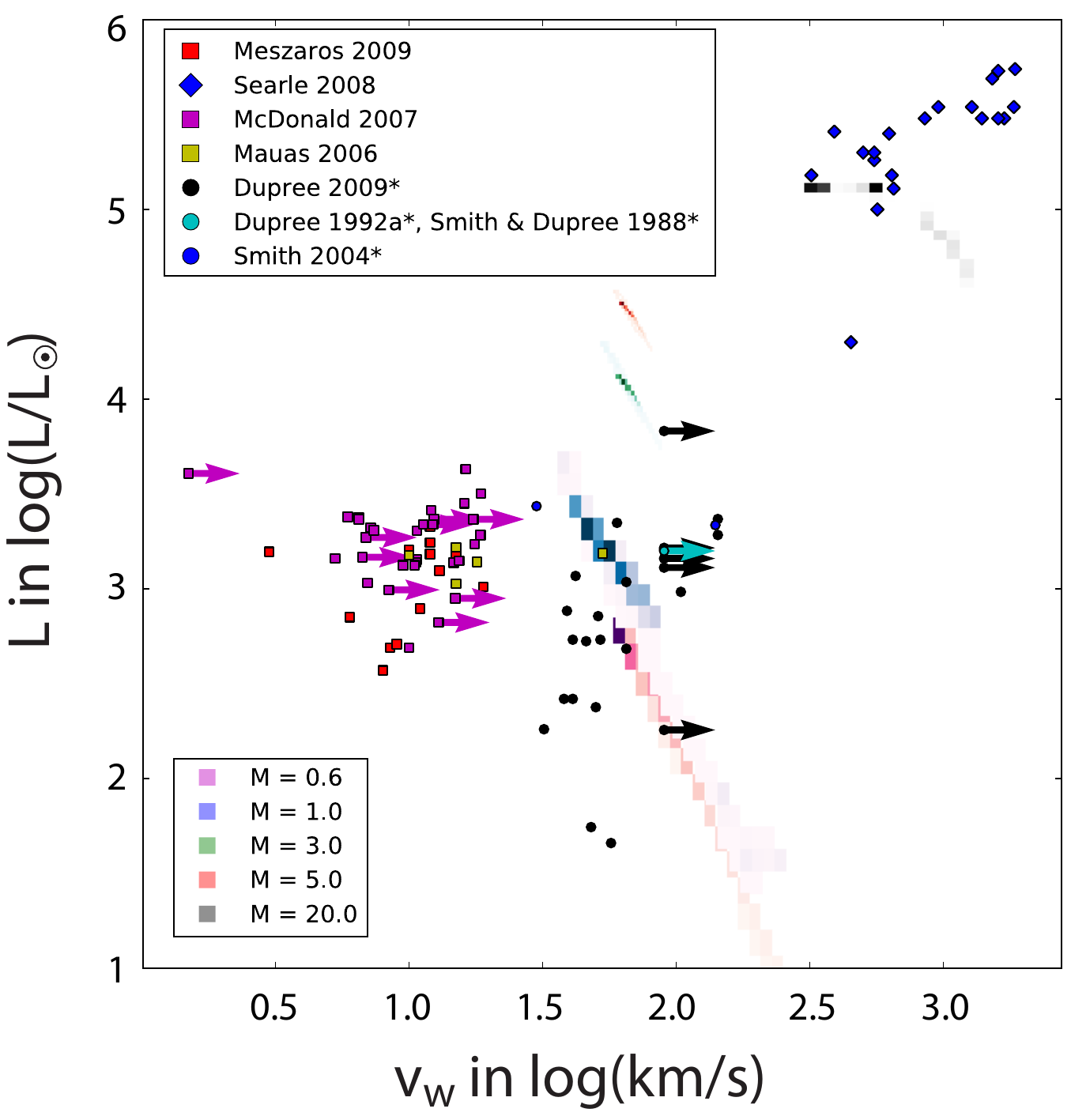}
\caption{Luminosity and wind velocities for AGB and RGB observations and MESA models are shown here with squares and circles while observations from high mass main sequence stars are shown with diamonds.  Two dimensional histograms plot the range of luminosities and wind velocities present in our MESA models, weighted by the mass lost in each bin, with the assumption that $v_w = v_{\rm esc}$.  For AGB and RGB stars, square points depict observations of outflows using H$\alpha$ and Ca~II triplet lines \citep{mauas2006,mcdonald2007,meszaros2009} while circles show observations made with the He~$I$~$\lambda 10830$ line \citep{smith1988,dupree1992a,dupree2009} which is generated higher in the atmospheres of giants \citep{dupree1992b}.  Observations noted with an asterix denote those with a measured $M_V$ which was converted to a luminosity for the sake of plotting.  The current masses of observed AGB and RGB stars are in the range $0.6-1.0 \, M_\odot$ \citep{smith1988,dupree1992a,mauas2006,mcdonald2007,meszaros2009,dupree2009}.   Masses for high mass main sequence stars range $10 \leq M/M_\odot \leq 20$ \citep{searle2008}.}
\label{fig:vwlum2}
\end{figure}

While the wind velocity can be approximated as the escape velocity, accurate within a factor of a few across a wide range of masses and life stages \citep{abbott1978,evans2004,schaerer1996,nyman1992,vass1993,loup1993,dupree1987,debes2006,badalyan1992}, in many clusters where $v_{\rm esc}$ for an individual star is comparable to the velocity dispersion of the cluster, a factor of two is the difference between significant mass retention in the cluster or large scale gas expulsion.
To test the accuracy of the assumption of $v_w \approx v_{\rm esc}$ we compare the predicted wind velocities from several MESA models with observations 
in Figure \ref{fig:vwlum2}.
H$\alpha$, Ca II H and K lines probe lower in the atmosphere and generally show outflows with velocities less than the escape velocity from the star at 
that depth (squares in Figure \ref{fig:vwlum2}), while the He~I~$10830$~\AA~ line is produced higher in the atmosphere and typically results in observed outward velocities comparable to the escape velocity of the star (circles in Figure \ref{fig:vwlum2}) \citep{dupree1992a,mauas2006,mcdonald2007,dupree2009,meszaros2009,smith2004}.
Given the lack of any observed inflow and the acceleration of the material from the lower layers of the AGB and RGB atmospheres it is probable that the velocity of the outflowing material continues to increase until it reaches the escape velocity from the star \citep{mauas2006}.
Thus, the assumption $v_w \approx v_{\rm esc}$ results in MESA stellar wind models which best reproduce the distribution of velocities lost from stars in unbound stellar winds as shown in Figure \ref{fig:vwlum2}.

 In addition to the mass lost from a star during its main sequence life time and late stages of evolution, material may be removed by close encounters with other stars \citep{pas2014}, or slower moving material may be stripped from the star as it moves through the intercluster medium \citep{wilkin1996}.  While stellar winds result in an injection of material at $v_w \approx v_{\rm esc}$, the last two mechanisms are possible avenues for incorporating slower moving (and thus, a lower injected kinetic energy material) into the intercluster medium if the lower atmospheric layers are stripped. 

One can estimate the effectiveness of ram pressure stripping on removing material from the lower atmosphere of RGB/AGB stars through the method outlined in \cite{wilkin1996} by calculating the stand-off radius of the bowshock as the star moves through the background medium:
\begin{equation}
R_{\rm rps} = 8.3 \times 10^5 \, R_\odot \left(\frac{\dot{M}}{10^{-8} \, M_\odot/{\rm yr}} \right) \left( \frac{v_w}{10 \, {\rm km s^{-1}}}\right) \left( \frac{n}{\rm cm^{-3}}\right) \left(\frac{v_\star}{30 \, {\rm km s^{-1}}} \right)
\label{eq:rps}
\end{equation}
For typical values of mass loss rate, $\dot{M} \sim 10^{-8} \, M_\odot/yr$, wind velocity, $v_w \sim 10 \, {\rm km s^{-1}}$, cluster velocity dispersion $v_\star \sim 30 \, {\rm km s^{-1}}$, and intercluster gas density $n \sim 1 \, {\rm cm s^{-1}}$, the interaction radius is much larger than the atmospheric extent of an RGB/AGB star, $R_\star \approx 100-200 \, R_\odot$ and therefore it is unexpected that ram pressure stripping plays a significant role in the majority of gas injection within a cluster.  However,  as the cluster gas density increases during high mass loss, low wind velocity phases this bowshock interaction radius may penetrate to the outer layers of RGB/AGB atmospheres - a intercluster gas density of $n = 10^{7} \, cm^{-3}$ results in a stand-off radius of $R_{\rm rps} \approx 200 \, R_\odot$.

Equation \ref{eq:rps} does not account for the non-constant nature of AGB mass loss which may result in periods of slow moving ejecta.  As noted in \cite{villaver2003,villaver2012} this periodic mass loss can lead to a large scale shell structure (1-2~pc) moving at slow velocities ($2-5 \, {\rm km s^{-1}}$) for AGB stars moving through the galactic ISM.  However, these simulations were conducted assuming slow moving winds emitted from isolated AGB stars, conditions quite different then found in most star clusters.

While ram pressure stripping may be effective in removing low velocity gas from stars in high {\it gas} density environments, ejection of stellar material during close star encounters could be prominent in high {\it stellar} density environments \citep{pas2014}.  
This may be an important route to include lower velocity wind material in the centers of the most dense globular clusters where interactions between stellar members is likely \citep{pooley2003,pas2014}.

In what follows we will ignore these possible secondary routes for the inclusion of stellar wind material at lower wind velocities and focus our attention on determining the effects of stellar winds alone on the mass and temperature of gas retained and ejected from star clusters.\\
\\

\subsection{Average Mass Loss and Wind Velocities as Model Inputs} \label{section:averages}

To finalize our mass and energy injection prescriptions it is necessary to derive an averaged mass loss rate and wind velocity during each phase of cluster evolution which includes contributions from the stars that significantly contribute to the mass and kinetic energy injection into the cluster.
The averaged prescription we present here has the added benefit of minimizing the effects of the poorly constrained, rapidly varying, final stages of a star's life \citep{pooley2006}.

\subsubsection{The Turn Off Approximation} \label{section:to}

To estimate the net mass loss and mean thermal velocities of the colliding winds,  we first follow the formalism developed by \cite{pooley2006} where the integrated wind kinetic energy and mass loss  of the stellar population's turn off star is used 
as  a proxy for the average wind velocity and mass loss rate:
\begin{equation}
\langle v_{\rm w, to}^2\rangle \approx {2\Delta E_K \over \Delta M}
\label{eq:vto}
\end{equation}
\begin{equation}
\langle \dot{M}_{\rm to} \rangle \approx {\Delta M \over \Delta t}.
\label{eq:mdotto}
\end{equation}
where $\dot{M}_{\rm to}(t)$ and $v_{\rm w, to}(t)$ are the mass loss rate and wind speed of turn off stars with a mass $M_\star$ at a particular cluster turn off time of $t$.
Here $\Delta E_K =  \frac{1}{2} \int_{t_0}^{t_1}\dot{M}_{\rm to} v_{\rm w, to}^2 dt$ and $\Delta M = \int_{t_0}^{t_1} \dot{M}_{\rm to} dt$ 
are the kinetic energy and mass loss input rates integrated over the lifetime  of the turn-off stars, 
$\Delta t = t_1 - t_0$, where $t_0$ is the zero age main sequence (ZAMS) and $t_1$ is the end of the AGB phase, white dwarf stage, or compact object creation.  This  provides a reasonable estimate for the overall wind mass and energy supply to the cluster although it fails to capture the variability of realistic stellar winds which are currently not well constrained \citep{marigo2012,wood2005,cohen2011}.
 
\subsubsection{The Population Averaged Approximation} \label{section:pa}
 
The effects of the additional input of mass and kinetic energy  by stars with $M_\star < M_{\star,{\rm to}}$, which are neglected in the formalism previously outlined, can be included by convolving  the above definitions of the average mass loss rate and wind velocity  with an initial mass function (IMF) and a star formation history.
 Following \cite{kroupa2013}, the average number of stars in a mass interval $[M_\star,M_\star+dM_\star]$ evolving between 
 $[t,t+dt]$ is given by  $dN = \zeta(M_\star,t) N_{\star} b(t) dM_\star dt$ where $\zeta(M_\star,t)$ is the IMF, which we assume  is accurately described by the \citet{kroupa2001}  IMF,  
 $b(t)$ is the normalized star formation history, and $N_{\star}$ is the total number of stars. For a non-evolving IMF, $\zeta(M_\star,t) = \zeta(M_\star)$, with a mass distribution extending 
 from masses $M_{\star,L}$ to $M_{\star,H}$, the normalized star formation history can be written as  $1/t_{\rm age}(M_{\star,L})\int_0^{t_{\rm age}(M_{\star,L})} b(t)dt = 1$, where 
$t_{\rm age}(M_{\star})$ is the  lifetime of a star of  a given $M_\star$. Examples of $b(t)$ include a constant star formation rate, $b(t)=1$,  or a coeval population, $b(t)= \delta(t -t_0)$, with all stars formed at $t_0$.

The average mass $\langle \Delta M(t_i)\rangle$  and kinetic energy  $\langle \Delta E_K(t_i) \rangle$  injected  into the cluster environment at  a time $t_i$,  by a population of stars with $M_\star \epsilon \;[M_{\star,L}, M_{\star,H}]$ where the lowest and highest mass in a population depends on both the age, $t_i$ and the birth rate of stars, with the birth rate regulated by $b(t)$, are then given by
 \begin{equation}
\langle \Delta M(t_i) \rangle = \int_{t_0}^{t_i} \int_{b(t_0)}^{b(t_i)}  b(t) \int_{M_{\rm \star, L}}^{M_{\rm \star, H}} 
\zeta(M_\star) \dot{M}(M_\star,t)  dM_\star db dt
\label{deltam}
\end{equation}
and
 \begin{equation}
 \begin{multlined}
\langle \Delta E_K(t_i) \rangle \\
= \frac{1}{2} \int_{t_0}^{t_i} \int_{b(t_0)}^{b(t_i)} b(t) \int_{M_{\rm \star, L}}^{M_{\rm \star, H}} 
\zeta(M_\star) \dot{M}(M_\star,t) v_w^2(M_\star,t) dM_\star db dt,
\label{deltae}
 \end{multlined}
\end{equation}
respectively.  Here, the lowest and highest masses existing at a given time depends on the current time and birth rate - $M_{\rm{\star,L}}(t_i,b)$, $M_{\rm{\star,H}}(t_i,b)$.
Stars that have lifetimes  $t_{\rm age}(M_\star) < t_i$ forsake the stellar population unless they were born during the most recent star formation time  interval $[t_i - t_{\rm age}(M_\star), t_i]$. For this reason,  $t_0$ is set to  $\max[0, t_i - t_{\rm age}(M_\star)]$. This formalism allows for equations (\ref{eq:vto}) and (\ref{eq:mdotto}) to be cast  into a more general form for the average mass loss rates and wind velocities at a specific stellar population life-time, $t_i$:
\begin{align}
 \langle \dot{M}(t_i) \rangle = {\langle \Delta M(t_i) \rangle \over  t_i} =  \frac{1}{t_i} \int_{M_{\rm \star, L}}^{M_{\rm \star, H}}  \zeta(M_\star) \dot{M}(M_\star,t)  dM_\star dt
 \label{eq:mdotpop}
\end{align}
and
\begin{equation}
\begin{multlined}
 \langle v_w^2(t_i) \rangle = {2\langle\Delta E_K(t_i)\rangle \over \langle\Delta M(t_i)\rangle}  \\
=  \langle \dot{M}(t_i)\rangle^{-1} \int_{M_{\rm \star, L}}^{M_{\rm \star, H}} \zeta(M_\star) \dot{M}(M_\star,t)  v_w^2(M_\star,t) dM_\star dt.
 \end{multlined}
\label{eq:vpop}
\end{equation}
In which we have assumed the stellar population initially residing in star clusters is coeval such that 
$b(t) = \delta(0)$, which provides an accurate description for clusters with $t_i \gtrsim 10$~Myrs.

\begin{figure}
\centering\includegraphics[width=0.46\textwidth]{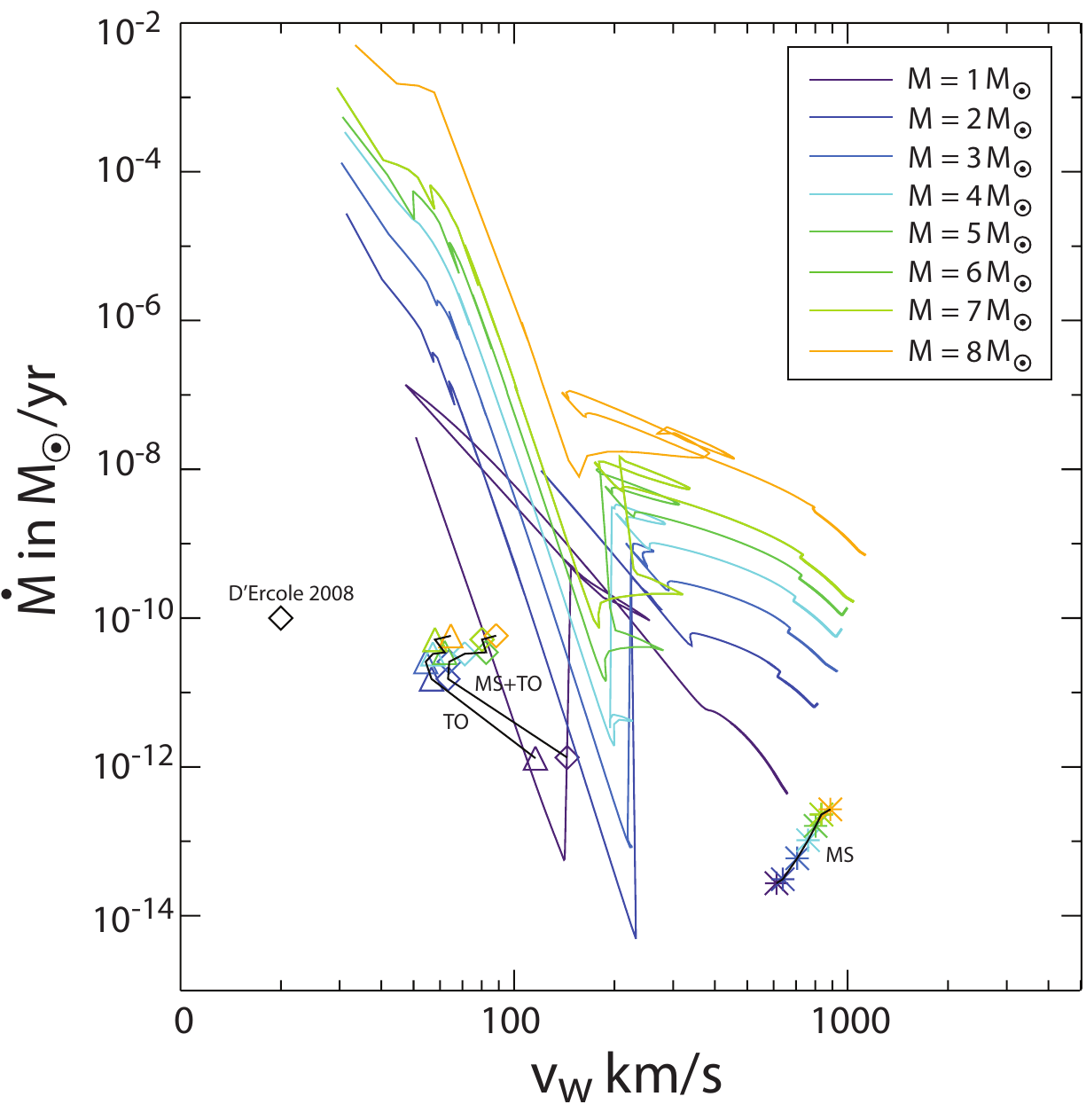}
\caption{ Mass loss rate and wind velocity for intermediate mass stars from MESA.  The average mass loss per star on the main sequence alone, using the turnoff mass prescription, and using the population averaged prescription are shown with colored stars (MS), triangles (TO) and diamonds (MS+TO), respectively.  For reference, the average mass loss per star from \citep{dercole2008} is shown by a black diamond at their AGB wind velocity of $20 \, {\rm km/s}$.}
\label{fig:fullmassloss}
\end{figure}

Figure \ref{fig:fullmassloss} compares mass loss rates and wind velocities of the the turn-off (TO) and population averaged (MS+TO) prescriptions with the instantaneous $\dot{M}-v_w$ curves for several MESA stellar evolution models.  While remaining similar in mass loss rates, the effects of the additional kinetic energy contribution in the population averaged prescription compared to the turn-off prescription are evident in the overall higher wind velocities in the TO curves of Figure \ref{fig:fullmassloss}.\\

In what follows we will explore the effects of substituting the mass loss prescriptions described in equations \ref{eq:mdotto} or \ref{eq:mdotpop} and the wind velocity prescriptions defined in equations \ref{eq:vto} or \ref{eq:vpop} on a cluster's overall mass loss and kinetic energy injection as defined in \ref{eq:mdotgen} and \ref{eq:kegen}, respectively.

\section{Stellar Wind Retention in Star Clusters} \label{section:results}

\begin{figure}
\centering\includegraphics[width=0.45\textwidth]{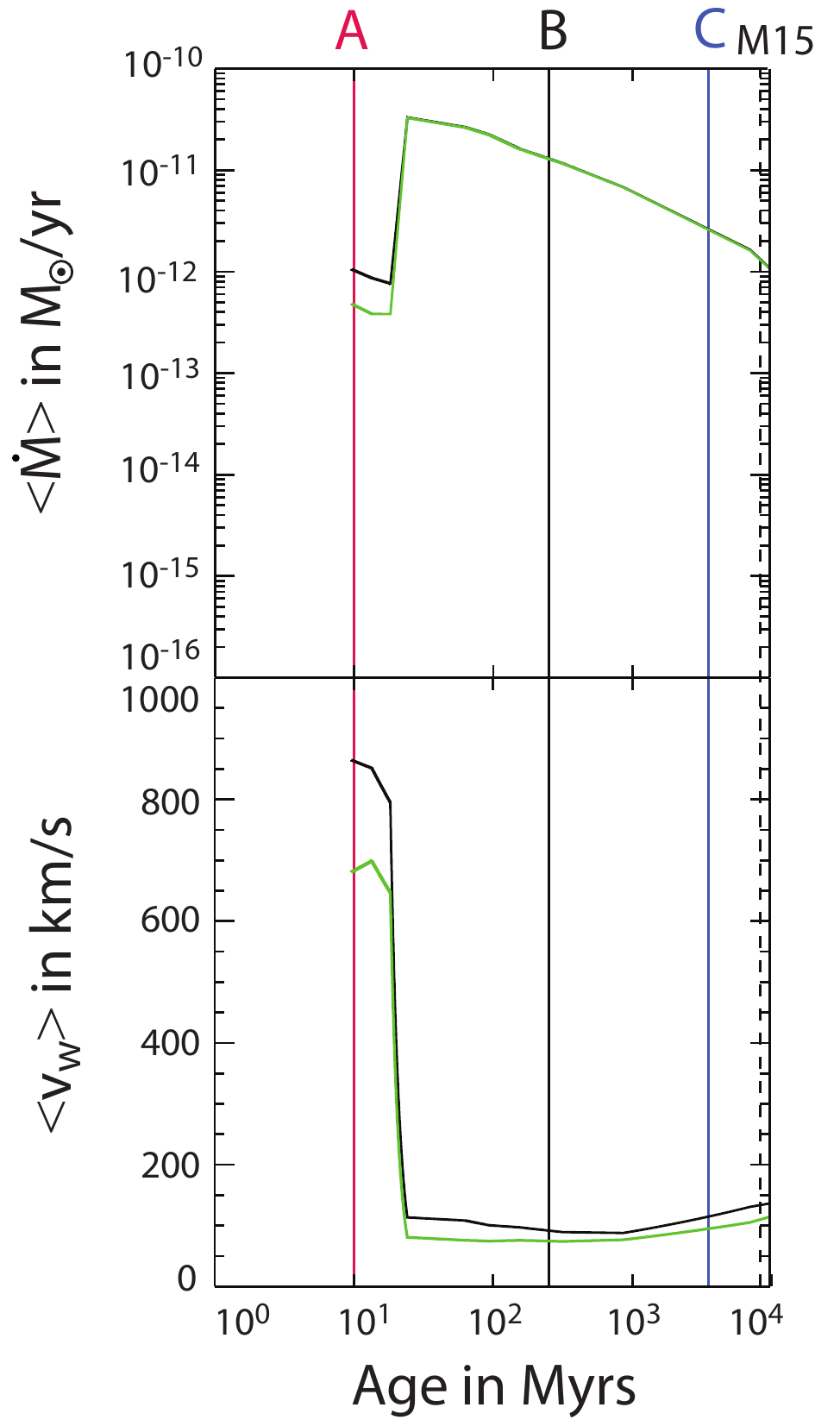}
\caption{Average cluster mass loss rates and wind velocities as a function 
of time for $Z = 1/10 Z_\odot$.
Green lines assume gas  dynamics are dominated by the wind properties of the turn off stars (\S \ref{section:to}), black lines 
show  the population averaged values (\S \ref{section:pa}).  Three representative times in the clusters age are denoted by the solid vertical lines.  
The current age of M15 is denoted by the vertical dashed line.  Note that while the turn off stars contribute the majority of the mass (top panel), the main sequence stars dominate the energy injection (bottom panel) as a result of their  higher effective wind velocities.}
\label{fig:fig1}
\end{figure}

The  mass injection properties, characterized here by the mass loss rates and wind velocities emanating from a coeval population of stars,  change dramatically  
as the population evolves.   Figure \ref{fig:fig1} shows the evolution of the average stellar wind parameters.  As the 
wind velocities and stellar mass loss rates change, so does the ability of a cluster potential 
to retain the shocked stellar wind  gas.  
 Figure \ref{fig:fig1} clearly illustrates  the differences  between efficient  (population averaged prescription which includes the kinetic energy injected by main sequence stars) and inefficient (turn off mass prescription which does not include the kinetic energy injected by main sequence stars) thermalization and mixing of stellar winds from stars of different masses. Note that the differences between the turn off mass and population averaged prescriptions manifest themselves  
predominately in the values of the population's averaged wind velocity as  main sequence stars do not contribute  to the total mass injection rate.

Figure \ref{fig:fig1} depicts the  wind properties as a function of time throughout the cluster's evolution.  If the cluster has its gas removed by, for example,  ram pressure stripping due to  interaction with external gas \citep{priestley2011} or by additional gas heating processes,  gas retention will commence without memory of  any previous episode of mass accumulation. If however, the cluster potential  is assumed to be non-evolving and isolated with no additional heating sources operating  other than 
stellar winds, there will be some residual gas and the gas retention properties at a particular age will depend on the mass and energy  injection history of the stellar members. We refer to these two extreme  scenarios as mass retention without memory and with memory, respectively.

\begin{figure*}
\centering\includegraphics[width=0.95\textwidth]{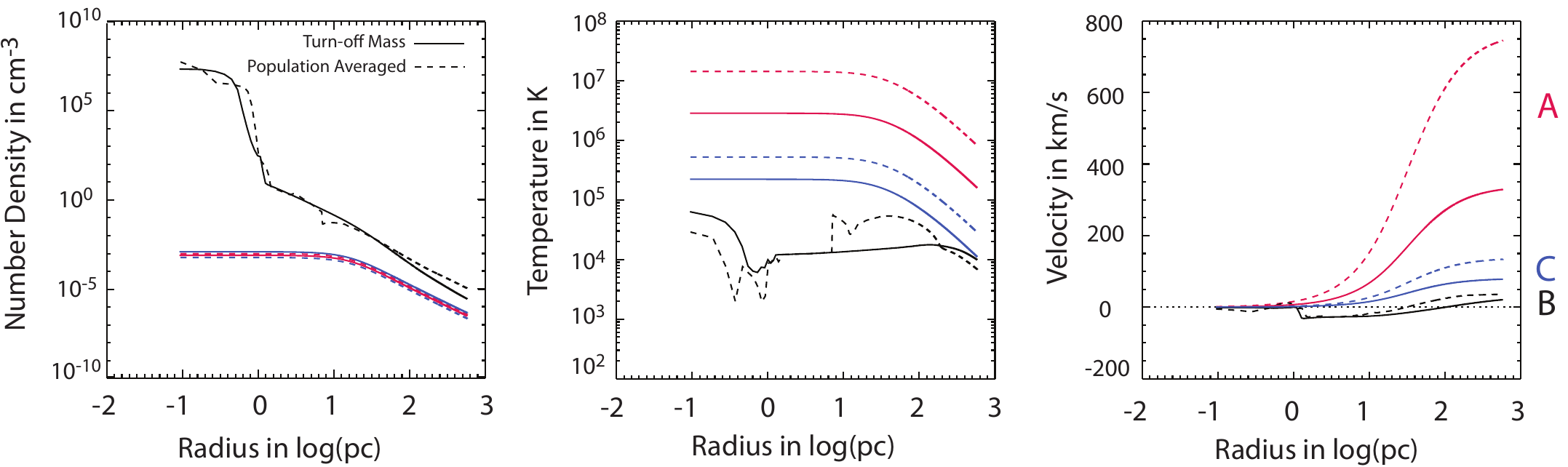}
\caption{Hydrodynamic profiles for the three representative times, A(red), B(black) and C(blue), as denoted in Figure \ref{fig:fig1}.  
These three plots are the density, temperature and velocity profiles at three 
representative times for a $M_{\rm c} = 10^7 \, M_\odot$, $\sigma_v = 30 \, {\rm km/s}$ model.   
Solid and dashed lines represent the turn off mass and population averaged prescriptions for models without memory, respectively.  }
\label{fig:fig2}
\end{figure*}

A young cluster without gas retention memory, corresponding to time {\it A} in  Figure \ref{fig:fig1}, with its high average wind velocity  
cannot retain gas effectively if  $\sigma_v\ll v_w$.  This is clearly seen  in Figure \ref{fig:fig2} which  shows the  properties of the shocked stellar wind gas confined to a cluster potential characterized by  $\sigma_v = 30 \, {\rm km/s}$ and $M_c = 10^7 \, M_\odot$ for both turn off mass and population averaged  prescriptions ({\it red} curves). The flow in this case was evolved for $t_i=t_A=23$~Myrs as the average mass loss rates and wind velocities do not change appreciably over this time period.  A young cluster  is thus only able to retain a small quantity of  high temperature gas in its inner  region ($M_{\rm acc}/M_{\rm c} \approx 10^{-7}$), while further away gas is blown out of the system in a wind. 

As the cluster members evolve
a dramatic decrease in the average wind velocity occurs due  to the dominant contribution  of the slow, dense AGB winds to the overall mass injection, corresponding to time {\it B} in   Figure \ref{fig:fig1}. For the star cluster, here assumed to be characterized by a non-evolving gravitational potential,  the AGB contribution  results in stellar wind injection parameters that favor significant  mass retention, as shown in Figure \ref{fig:fig1} ({\it black} curves).  As the stellar wind gas shocks and   cools, mass is efficiently accumulated ($M_{\rm acc}/M_{\rm c} \approx$0.04) until the central region becomes Jeans unstable, thus triggering star formation. In this case, the  gas flow within the cluster is  evolved until $t = t_{\rm sf}$, defined here as the sound crossing time within the  (adequately  resolved) Jeans unstable region. Here $t_{\rm sf} = $ 1200~Myrs and 1700~Myrs  for the turn-off mass  prescription and population averaged approximation, respectively.

As the cluster ages, remaining stars no longer pass through an extended thermally pulsing AGB phase.   This results in a increase in 
average wind velocity, and decrease in average mass loss rates (by a factor of $\sim$1.3 and approximately an order of magnitude over $\sim$2~Gyrs, respectively)  as seen at time {\it C}.
As a result, the cluster potential  is unable to effectively retain the emanating gas and the remaining wind material flows out of the cluster almost unrestrained ($M_{\rm acc}/M_{\rm c} \approx10^{-7}$).

\begin{figure*}
\centering\includegraphics[width=0.95\textwidth]{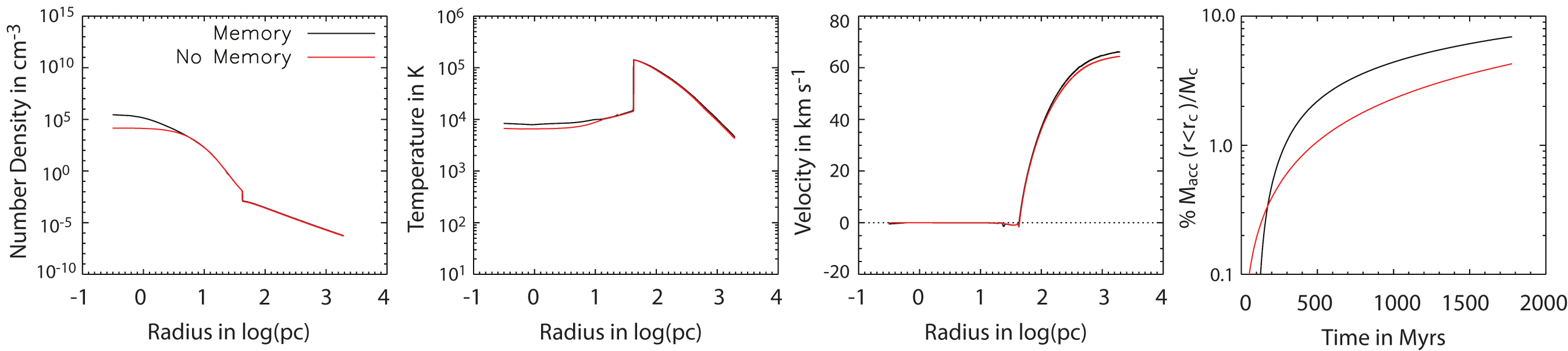}
\caption{Hydrodynamic profiles and gas accumulation for a $t_i = 2000 \, {\rm Myrs}$ cluster with  $M_{\rm c} = 10^7 \, M_\odot$
and $\sigma_v = 27 \, {\rm km/ s}$.  
When the cluster is allowed to retain residual gas throughout its evolution (black line) it can accumulate 
significantly more mass ($6.9\%$ instead of $4.3\%$) by the time star formation is triggered than when the cluster is assumed to have no retention memory.}
\label{fig:memhydroprofiles}
\end{figure*}

The amount of gas in clusters in which stellar wind material is efficiently retained, as in time {\it B} of Figure \ref{fig:fig1}, can increase significantly if, prior to this time, the cluster is not stripped of gas prior to this time by  additional internal  heating sources or by external gas removal. In this case the cluster can retain the memory of previous gas accumulation.
Figure \ref{fig:memhydroprofiles} compares the hydrodynamical profiles and central gas mass accumulation histories of two clusters in which mass retention is assumed to take place with and without memory.  In both cases, star formation is triggered albeit at different times and involving different amounts of accumulated cold gas. The larger central gas density in the model with memory and residual gas results in a few percent increase in cold gas made available for a second generation of star formation - $M_{\rm acc}/M_{\rm c}$ increases from $4.3\%$ to $6.9\%$ when the mass retention takes place with memory.

\begin{figure*}
\centering\includegraphics[width=0.95\textwidth]{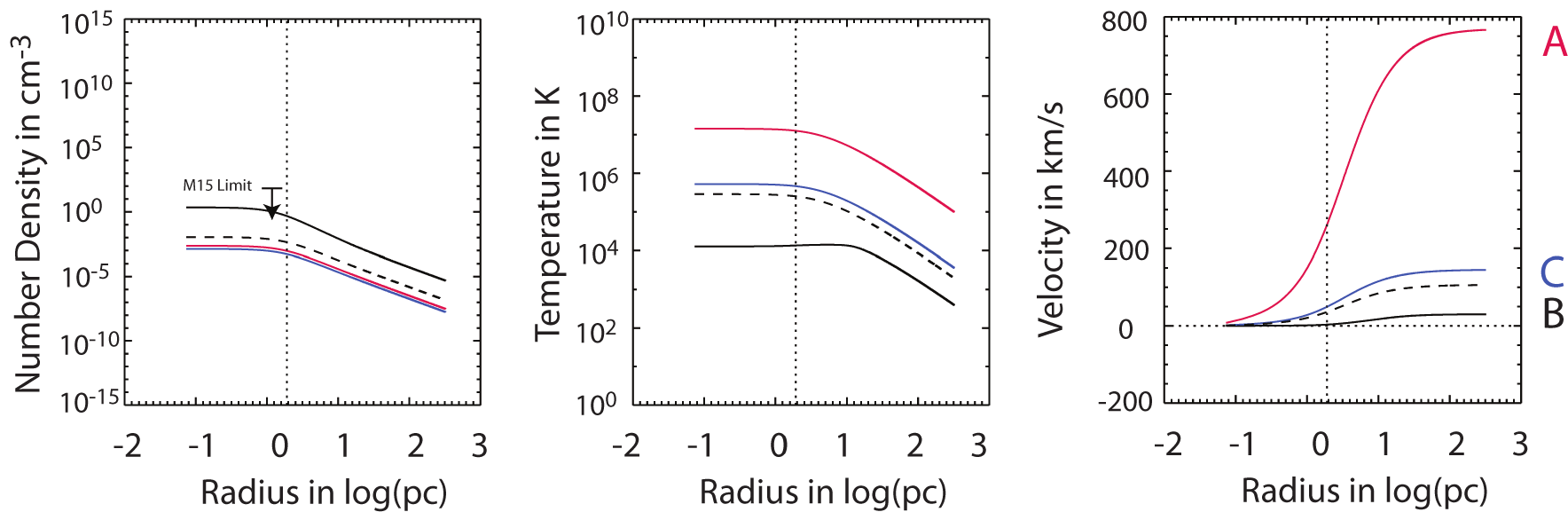}
\caption{Hydrodynamic profiles for M15 with $M_{\rm c} = 4.5 \times 10^5 \, M_\odot$
and $\sigma_v = 14 \, {\rm km/s}$  \citep{mcnamara,gerssen} at the three representative times in Figure \ref{fig:fig1}, including the predicted profiles for 
its current age (dashed lines).   Here, the population averaged prescription is used and  we assume  the cluster has no gas retention memory.}
\label{fig:fig3}
\end{figure*}

The ability for a star cluster to retain wind ejecta depends not only on the evolutionary stage of its members but also on their spatial distribution. To illustrate this, in Figure \ref{fig:fig3} we show the  properties of the shocked stellar wind gas confined to a shallow potential characterized  by 
$\sigma_v = 14 \, {\rm km/s}$ and $M_{\rm c} = 4.5 \times 10^5 \, M_\odot$ (parameters  thought to accurately represent the current  stellar mass distribution in the globular cluster M15 \citep{mcnamara,gerssen}).  A shallow gravitational potential is thus unable to retain a significant amount of stellar wind material, even during the AGB phase.   This is observed to also be the case in simulations  in which the cluster is assumed to have gas retention memory. 

\begin{figure*}
\centering\includegraphics[width=0.95\textwidth]{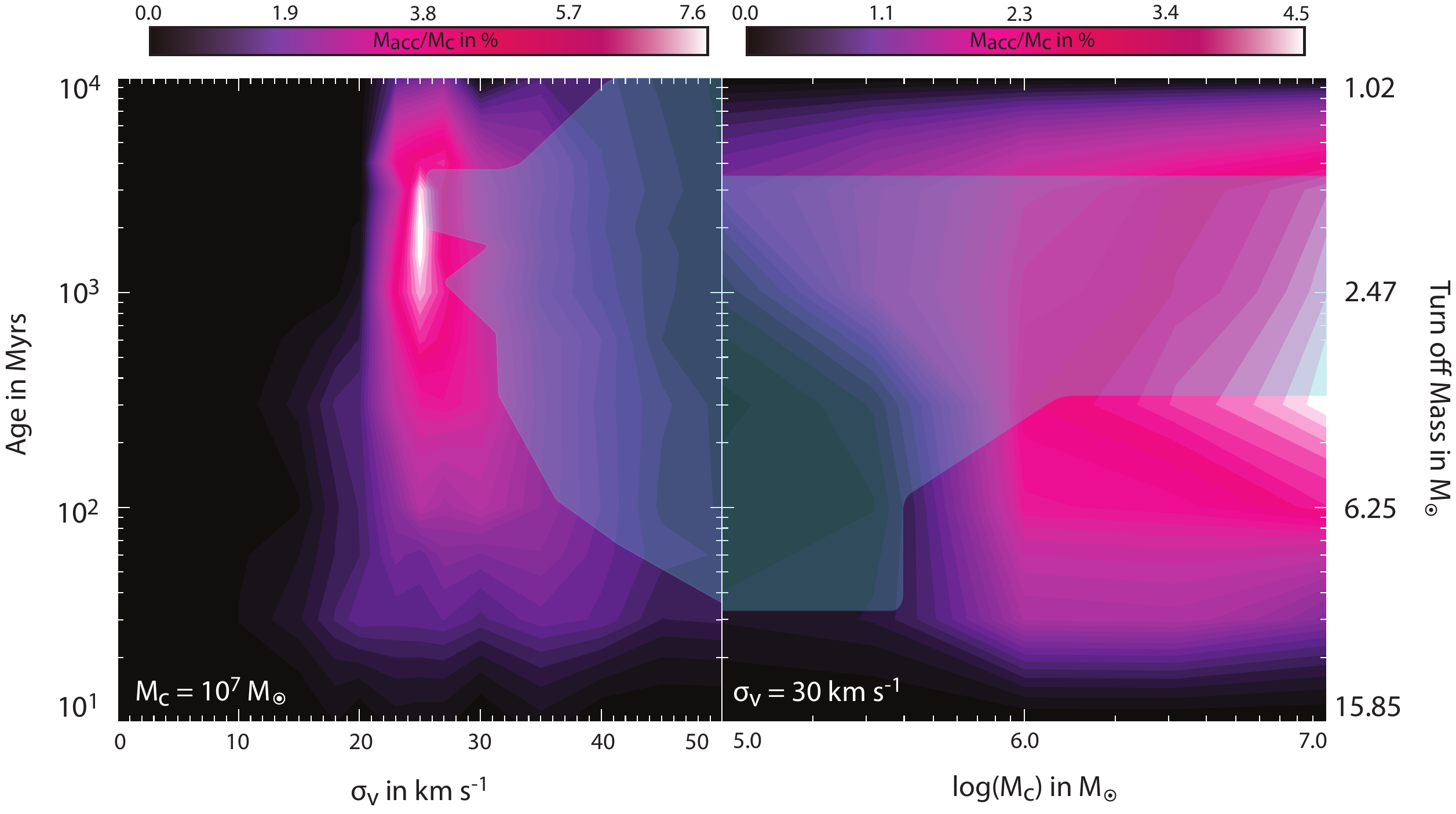}
\caption{Mass accumulation, $M_{\rm acc}$, as a function of the potential parameters $M_{\rm c}$ and $\sigma$ for the 
turn off mass prescription.   The blue shaded regions delimit the boundaries for which our models collapse and trigger star formation.
{\textit{Left:}} Contours of $M_{\rm acc}/M_{\rm c}$ for a fixed cluster core 
mass of $M_{\rm c} = 10^7 \, M_\odot$ and a varying velocity dispersion. 
 {\textit{Right:}} Contours 
of $M_{\rm acc}/M_{\rm c}$ for a fixed velocity dispersion of $\sigma_v = 30 \, {\rm km/s}$ and 
varying cluster mass.  Here we assume the cluster has no gas retention memory.}
\label{fig:fig4}
\end{figure*}

The dramatic difference observed in Figures \ref{fig:fig2} - \ref{fig:fig3} between the stellar wind mass accumulated  in  cluster potentials of varying properties  motivates our study to compare  results  obtained with different velocity dispersions and cluster masses for clusters with and without gas retention memory. To facilitate  comparisons, we first systematically vary the  cluster velocity dispersion 
for a fixed  total mass $M_{\rm c} = 10^7 \, M_\odot$, assuming the turn off mass wind velocity and mass loss rate are 
representative of  the entire population and the cluster has no gas retention memory.
 The left panel of Figure \ref{fig:fig4} gives the amount of accumulated stellar wind mass in units of $M_c$
for a range of cluster velocity dispersions and different evolutionary stages ($t_i$) of the stellar members. 
By looking at the shaded regions in Figure \ref{fig:fig4}, the reader can identify   the cluster velocity dispersion and mass combinations for which favorable conditions for star formation are satisfied at a given cluster age provided there is no memory of gas retention.

Cluster potentials with $\sigma_v \lesssim 20 \, {\rm km/s}$ are not effective at retaining gas 
in their cores, and gas is blown out of  the cluster at all times. 
As the velocity dispersion increases above  $30 \, {\rm km/s}$, the cluster core is able to retain the shocked and efficiently  cooled stellar wind gas, making  the  central region Jeans unstable before (i.e. $t_{\rm sf} \ll t_i$)  a significant amount of mass is  accumulated  into 
the cluster. The largest fraction of mass  retained  takes place  for star clusters  with $\sigma_v \approx 25 \, {\rm km/s}$. In this case, the potential  is steep enough to retain the shocked and subsequently  cooled wind material, but  shallow enough to force the  cold gas to  collapse only until a significant amount of mass has been accumulated.
The contours in the left panel of Figure \ref{fig:fig4} show that potentials with  $\sigma_v \gtrsim 30 \, {\rm km/s}$  may 
endure multiple episodes of star formation albeit  involving less retained gas, while  
potentials with  $25\, {\rm km/s} \lesssim \sigma_v < 30 \, {\rm km/s}$  can have only short but more intense star formation  periods. 
However, because we do not treat star formation explicitly, future detailed simulations are needed to address the possibility of recurrent star formation in these systems.
The largest mass accumulation for a single star formation episode  occurs for $\sigma_v \approx  27 \, {\rm km/s}$ between the ages of $1-10$~Gyrs, 
with a mass retention fraction of $M_{\rm acc}/M_c \approx 6.4\%$.

The extension of the star formation region in Figure \ref{fig:fig4} depends dramatically on the velocity dispersion of the cluster.  While our optimal  single star formation episode takes place  between cluster ages of $1-10$~Gyrs, significant levels of star formation can occur  at earlier times for higher velocity dispersions - clusters with younger ages, $\approx 200-400$~Myrs, can retain approximately 3\% of the original cluster mass provided $\sigma_v \gtrsim 30 \, {\rm km/s}$.

For consistency with other works we have repeated our analysis shown in right panel of Figure \ref{fig:fig4} including the effects of the cluster velocity dispersion term which is often added to the energy relation derived in equation \ref{eq:energy}.  For brevity, the contour map is not included here but we note that we find that in compact clusters this extra energy injection term slightly delays star formation until enough mass has been accumulated in the central regions to cool efficiently and subsequently  trigger star formation.  However, for less compact clusters ($\sigma_v \lesssim 35 \, {\rm km/s}$) this additional heating is not important.  

The right panel of Figure \ref{fig:fig4}, in conjunction with the left panel, allows us to estimate  the amount of stellar wind material retained by a star cluster of a given age, mass and velocity dispersion. 
While our models do not span the full range of $M_{\rm c} - \sigma_v$ combinations,  some generalizations can be made  from the results presented in Figure \ref{fig:fig4}.
The conclusions drawn are based upon the assumption that the cluster potential properties, whose evolution causes are poorly known, remain relatively unaltered for $\min [t_i,t_{\rm sf}]$ and that the cluster has no gas retention memory.
The right panel of Figure  \ref{fig:fig4} is self-explanatory - heavier star clusters at a fixed velocity dispersion retain more gas. The mass retention fraction  is observed to  increase slightly  with augmenting  cluster mass as a result of less efficient cooling, which in turn delays the triggering of star formation as mass continues to accumulate.
 Depending on how the cluster evolves, the assumption of a static potential used in Figure \ref{fig:fig4} may brake down, however the effects of a time varying potential can be estimated by altering the trajectory of a cluster in the [$\sigma_v$, time]   or  [$M_{\rm c}$, time] plane.
 
\begin{figure}
\centering\includegraphics[width=0.48\textwidth]{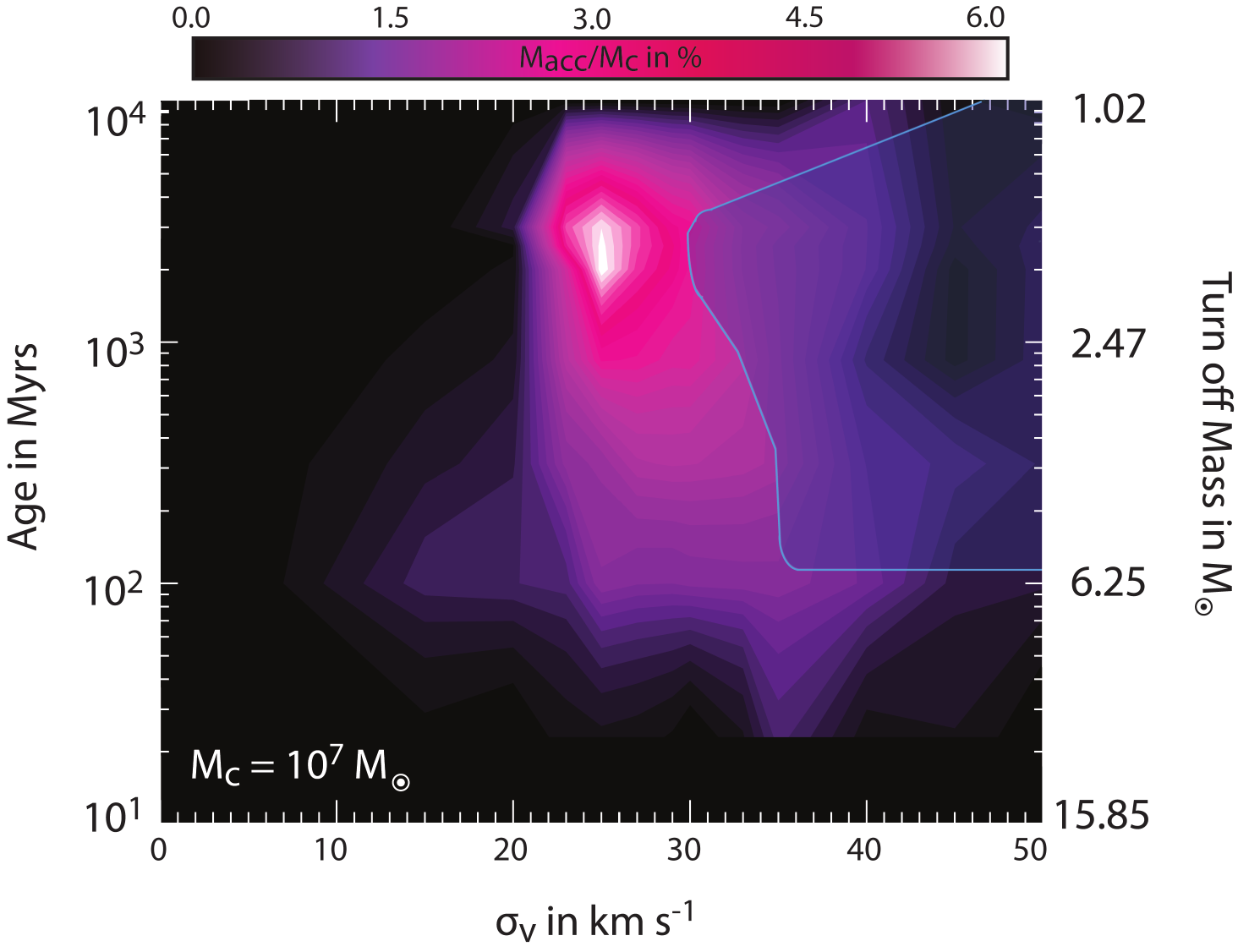}
\caption{Mass accumulation, $M_{\rm acc}$, as a function of  $\sigma_{\rm v}$ for a fixed cluster core 
mass of $M_{\rm c} = 10^7 \, M_\odot$, calculated using  the population averaged 
prescription.  The blue shaded region delimits the boundaries for which our models collapse and trigger star formation. Here we assume the cluster has no gas retention memory.  The additional thermalized high velocity winds from main sequence stars add appreciable heat to the central regions of the cluster, thus inhibiting star formation at lower cluster velocity dispersions.}
\label{fig:fig5}
\end{figure}

Using the average wind velocity and mass loss rate  of the turn off mass star alone to represent the properties of the entire population underestimates  the kinetic energy 
input arising  from the more numerous, lighter stars (Figure \ref{fig:fig1}). This  gives  an  optimistic  value  of the  wind retention and star formation efficiencies.
The effects of using  the total kinetic energy input from the star cluster, including the contribution from the main sequence stars, to calculate effective mass retention  are shown in  Figure \ref{fig:fig5}.  Direct comparison  with Figure \ref{fig:fig4} shows that although  the overall mass accumulation is rather similar, in the population-averaged prescription  star formation is triggered over a narrower range of velocity 
dispersions.  Not only are higher velocity dispersions required to produce efficient star forming models, but the additional kinetic energy  inhibits star formation in the largest mass accumulation regions.  This is because  the larger wind velocities keep the gas at higher  temperatures, thus quenching  star formation.  
Previous  studies used the instantaneous AGB wind velocity as proxy for the total kinetic energy being injected into the cluster.  Our results suggest that the inclusion of the kinetic energy provided by  the main sequence stellar winds  can dramatically alter the gas dynamics in 
these systems.  However, the ability of stellar winds from  different populations to mix and effectively thermalize remains uncertain and a clear understanding of their combined effects  will  require detailed multidimensional simulations, which are currently beyond the scope of this paper.

\begin{figure}
\centering\includegraphics[width=0.49\textwidth]{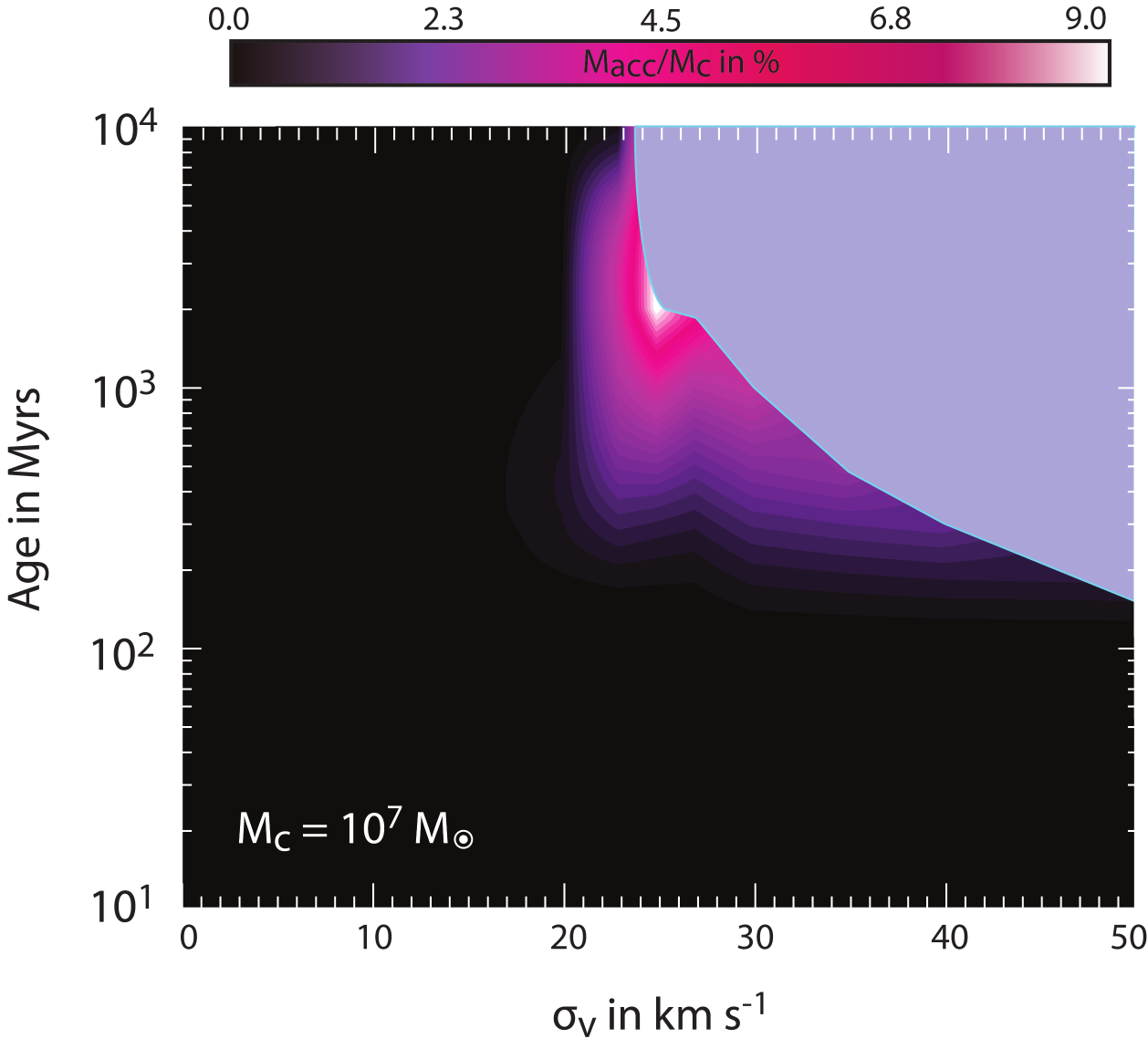}
\caption{Mass accumulation, $M_{\rm acc}$, as a function of  $\sigma_{\rm v}$ for a fixed cluster core 
mass of $M_{\rm c} = 10^7 \, M_\odot$, calculated using  the population averaged 
prescription.  The blue shaded region delimits the boundaries for which our models collapse and trigger star formation.  Here we assume the cluster has gas retention memory. After star formation is triggered we expect the subsequent gas accumulation after supernova energy injection to initially proceed as in Figure~\ref{fig:fig4}.  Here, the additional mass input increases the cooling capacity of the cluster gas when compared to that in Figure~\ref{fig:fig5} and, as found previously,   the star forming contours overlap with the largest central mass accumulation regions, as in Figure~\ref{fig:fig4}. }
\label{fig:memcontour}
\end{figure}

The effects of additional kinetic energy injection by main sequence stars in star clusters can be mitigated by the larger densities and thus enhanced  cooling rates present in models where gas retention is allowed to proceed unimpeded  throughout the cluster's evolution.  Figure \ref{fig:memcontour} depicts the effects of gas retention with memory on the central mass accumulations and star formation histories of a $M_c = 10^7 \, M_\odot$ cluster for a variety of velocity dispersions.  Note that our rudimentary  treatment of star formation prevents us from following the gas evolution once star formation has been triggered, resulting in an incomplete coverage of the $M_c$-$\sigma_v$ plane in Figure \ref{fig:memcontour}. In Figure \ref{fig:memcontour}, the largest mass accumulation coincides with the lower $\sigma_v$ bound of the star formation region.  The continuous accumulation of gas also produces larger central gas masses, with a gas mass of up to $9\%$ of the cluster's mass retained.  A  comparison between the mass accumulation contours in Figure \ref{fig:memcontour} with those in the left panel of Figure \ref{fig:fig4} should provide the reader with some understanding of the importance of gas retention memory  although such comparison should be done with care, as the assumption of a non-evolving potential in Figure \ref{fig:memcontour} is not necessarily a good approximation over a large expanse of time.

In estimating the effects of star formation episodes in Figures \ref{fig:fig4} - \ref{fig:memcontour}, we assume that such events do not effect the average stellar mass loss parameters dramatically.  This approximation will hold true provided that  number  of stars  form during these episodes is small, as suggested by our work. In such cases, the shaded regions depicted in Figures \ref{fig:fig4} - \ref{fig:memcontour} provide valid  constraints on  the star formation ages and masses expected from stellar wind  retention.

\section{Further Considerations} \label{section:discussion}

Throughout this work, we have used simplified  models to determine how effective  a  star cluster of a particular age  is at retaining  gas emanating from its  stellar members, under the assumption that its potential remains unaltered during  the simulated phase. We have further assumed a single metallicity for all clusters and have disregarded any heating sources besides the stellar winds themselves.  In this section we relax both of these assumptions.

\subsection{Metallicity}

Not all clusters will have stars which expel wind material at $1/10 \, Z_\odot$.
Cluster to cluster variations in light element abundances are commonly  observed in Galactic globular clusters  \citep{beasley,caldwell,gratton2004,carretta2010c,cohen2005,piotto2007}, with spreads in heavier elements present in the most massive Galactic globular clusters \citep{marino2009,carretta2009a,ferraro2009,carretta2010a,carretta2010c}.   Additionally, Galactic globular clusters have metallicities which vary from cluster to cluster -  [Fe/H] ranges from approximately $ -0.5$ to $ -2.5$ \citep{harris1996}, and there is evidence these metallicity variations extend to other galaxy and globular cluster systems \citep{usher2015}. 
In addition, there are some observational hints at the existence of complex abundance patterns in younger clusters as well \citep{li2016,martell2013}.  While it is not clear how robust this effect is within all young and intermediate age clusters \citep{mucc2008,mucc2014}, there are many examples of YMCs and IACs with metallicities not equal to that in our assumed models, $Z = 1/10 \, Z_\odot$ \citep{mucc2008}.

These variations may cause changes in the mass loss histories  of the individual stellar members and the cooling properties of the shocked gas. As the stellar mass loss prescription are mostly independent of metallicity during the evolutionary time periods that are conducive to star formation ($0.1 \, {\rm Gyrs} \lesssim t_{i} \lesssim 100 \, {\rm Gyrs}$) \citep{reimers1975,blocker}, we account for the effects of  varying metallicity solely in the  cooling function. We further assume all changes are factors of solar metallicity - only an approximation when enrichment follows the abundance variations seen in present day globular clusters.  

\begin{figure*}
\centering\includegraphics[width=0.95\textwidth]{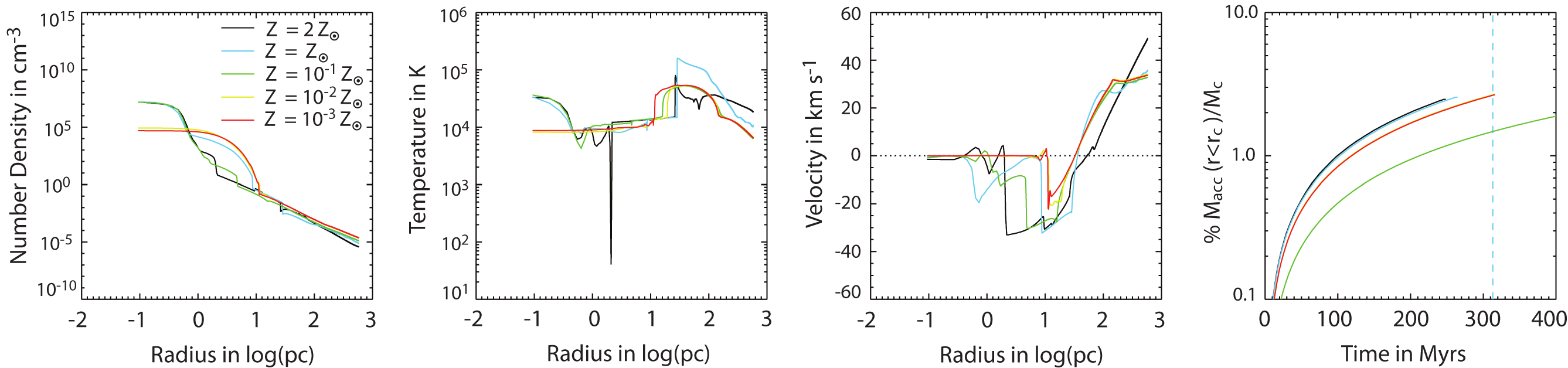}
\caption{Gas properties for $M_c = 10^7 \, M_\odot$, $\sigma_v = 30 \, {\rm km/s}$ cluster 
at $t_{i} = 313 \, {\rm Myrs}$ for different metallicities.    Here, we use the population averaged values of the wind parameters and assume the cluster has no gas retention memory.}
\label{fig:fig7}
\end{figure*}

Figure \ref{fig:fig7} shows the effects of changing metallicity for a young, massive and compact star forming cluster, described here by 
$M_c = 10^7 \, M_\odot$, $\sigma_v = 50 \, {\rm km/s}$, 
and $t_{i} = 212 \, {\rm Myrs}$.   
For lower metallicities, the cooling becomes less efficient and  more material  is allowed to flow into the center of the cluster before catastrophic cooling occurs. This enables relatively more massive star forming episodes to be triggered.   Interestingly,  when metallicity is lower than $Z < 10^{-1}$, the cooling  becomes weak  enough to prevent  catastrophic cooling at times  $\leq t_{i}$.  
These results suggest that the range of cluster parameters  over which large central densities will persist before catastrophic cooling takes place (Figures \ref{fig:fig4} - \ref{fig:memcontour})  will depend  on the metallicity of the emanating stellar  winds, 
though,  as illustrated in Figure \ref{fig:fig7}, the differences are not marked.

\subsection{Intercluster Heating Sources}

In addition to altering the cooling curves, the inclusion of additional cluster heating sources may prevent effective gas retention in our simulations.  We address this 
problem here by artificially increasing the energy input rate: 
$q_{\epsilon,{\rm new}} = (1+H)q_\epsilon = (1+H)\frac{1}{2} q_m(r)v_w^2$. Under this assumption, the additional  heating sources follow the potential's stellar distribution.  Such a heating source could be, for example, the result of including the extra velocity dispersion term to the energy relation shown in equation \ref{eq:energy} by setting $H\approx1$.

\begin{figure*}
\centering\includegraphics[width=0.95\textwidth]{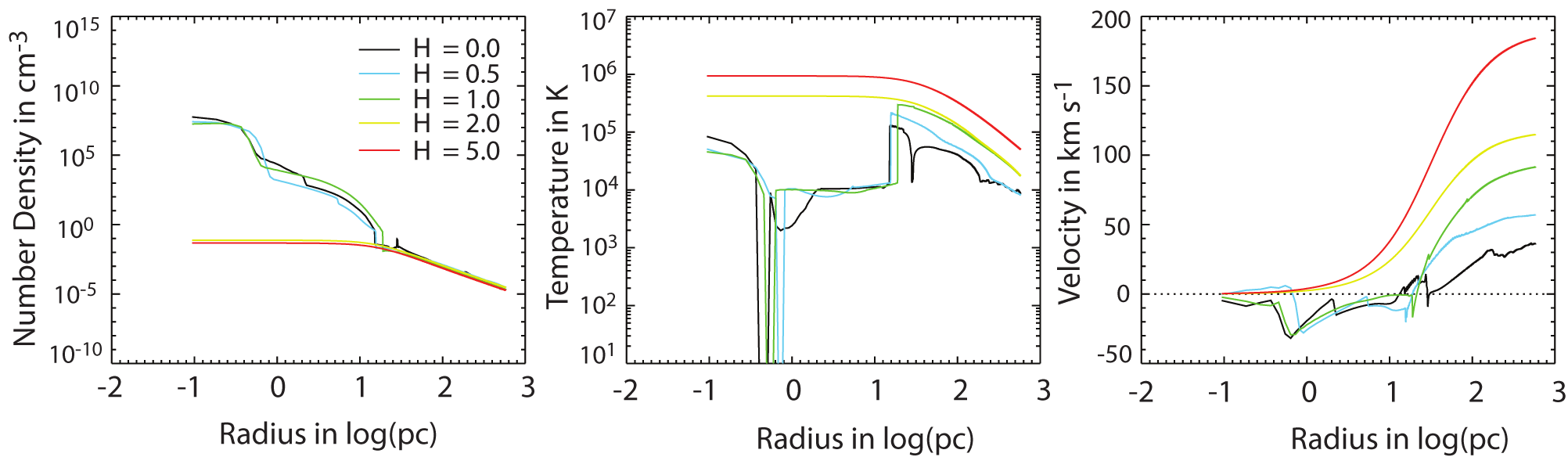}
\caption{{\textit{Left 3 Plots:}} Gas properties for $M_c = 10^7 \, M_\odot$, $\sigma_v = 30 \, {\rm km/s}$ cluster 
at $t_{i} = 1860 \, {\rm Myrs}$ with added heat in the form of thermal energy: 
$e_{int,new} = (1 + H)e_{int,old}$.   Here, we use the population averaged values of the wind parameters and assume the cluster has no gas retention memory.}
\label{fig:heating}
\end{figure*}

Figure \ref{fig:heating} shows the effects of the additional  heat input in one of our otherwise star forming simulations.  For $H < 2.0$, the gas in the simulation still collapses, triggering star formation.  For larger  values, on the other hand, the cluster is unable to effectively retain the gas  and, as a result, star formation never ensues.  By integrating $q_{\epsilon,{\rm new}}$ over the cluster's core for $H=2.0$, we derive the total energy input rate required to overturn the central mass build up, which for this simulation is about  $10^{35} \, {\rm erg s^{-1}}$. 

In many  cases, the additional energy injection sources might not follow the stellar distribution.  As an example, let's compare the heat distribution  expected from accreting neutron stars under the assumption that the accretion feedback  is proportional to the Bondi accretion rate:  
$q_{\epsilon,{\rm ns}} \propto \dot{M} \propto \rho(r) T(r)^{-3/2}$ \citep{bondi}.   Using  the volume-averaged density and 
temperature in the 
cluster core, $\bar{\rho}  \approx 10^{-22} \, {\rm g\; cm^{-3}}$ and $\bar{T}  \approx 10^4$K,  we derive  the average  luminosity of a single, accreting neutron star: $L_{NS} \approx 10^{33} \, 
{\rm ergs\; s^{-1}}$.  This implies that  $\gtrsim$ 100 accreting neutron stars are required to reside in the cluster's core  in order to significantly offset  its cooling  properties. However, to accurately test this phenomena a multi-dimensional  approach would be required as feedback  would not 
necessarily act as a simple heating prescription.

\section{Applications to Various Star Clusters} \label{section:applications}

We have thus far kept our discussion of stellar wind retention in star clusters generalized to star clusters with different combinations of ages, masses and velocity dispersions.  In what follows we discuss the implications of gas retention on three specific groups of star clusters - Young Massive Clusters, Intermediate Age Clusters, and Globular Clusters - and provide limits on how stellar wind retention can effect their star formation histories.

\subsection{Young Massive Clusters: Age $\lesssim$1~Gyr}

Young, massive star clusters (YMCs) are ubiquitous in the nearby Small and Large Magellanic Clouds \citep{goud2014} and are also found in recent ($\lesssim$1~Gyr) mergers like the Antennae galaxies \citep{whitmore2007}.  As one of the canditates for proto-Globular Clusters, the gas content and star formation histories of these objects are a subject of much interest \citep[e.g.][]{zwart2010}.

\subsubsection{Current Gas Content} \label{section:YMCcurrent}

 During the time span of the typical ages of YMCs (10s-100s~Myrs) Figures \ref{fig:fig4} - \ref{fig:memcontour} show little mass retention and star formation in all but the most compact stellar clusters.
 This results from the fast winds from main sequence O and B stars as shown in Figure \ref{fig:fig1} for ages $\lesssim 20$~Myrs, and the lack of sufficient gas accumulation time to initiate runaway cooling and collapse for clusters with $20 \lesssim {\rm age/Myrs} \lesssim 500$.
 For clusters with large velocity dispersions, $\sigma_v \gtrsim 35 \, {\rm km s^{-1}}$, while some gas mass is retained and star formation is triggered, less than $\approx 2$\% of the original cluster's mass is available for star formation.

 In such clusters we expect several additional mechanisms to add significantly to the energy injection rate in the intercluster gas.
 As shown in \cite{calura2015} SNe can be an effective avenue to assist in the removal of gas from stellar clusters, though in some systems SNe alone may not be able to fully remove gas from young clusters \citep{krause2013}.
 In addition, accretion onto compact objects may be able to clear gas within YMCs on time scales as small as 10~Myrs \citep{leigh2013a}.
 Finally, Lyman-Werner flux from massive stars further inhibits star formation by dissassociating molecular hydrogen in these young systems \citep{ten1986,conroy2011b,krause2013}.
 Therefore, we conclude that our estimates of gas retention on the order of a few percent are upper limits for the total mass retained from stellar wind ejecta within YMCs.

 Such a small amount of gas retention is broadly consistent with both observations and more complex simulations.
 Recent observations which show little gas in all but the most compact clusters \citep{bastian2014a,cabrera2015,krui2015,longmore2015}, though observations at high redshift are challenging \citep{longmore2015}.
 Our results are also consistent with analytic and three dimensional simulations of  which find that the majority of gas mass is removed by $1-14$~Myrs \citep{calura2015,krui2015}.

\subsubsection{Previous and Ongoing Star Formation} \label{section:YMCongoing}

 Our results of little to no star formation in Figures \ref{fig:fig4} - \ref{fig:memcontour} over the several hundreds of Myrs timespan are broadly consistent with the results of \cite{bastian2013a} which show no ongoing star formation in 130 YMCs with masses ranging from $10^4 < M/M_\odot < 10^8$ and ages $10 < {\rm age/Myrs} < 1000$ and \cite{mart2018} who only find MSPs in clusters with ages $\gtrsim 2$~Gyr, though our level of predicted star formation may be too low to be detectable in the majority of YMCs \citep{peacock2013}.
  Our limit of $\sigma_v \gtrsim 35 \, {\rm km s^{-1}}$ is a slightly higher limit for velocity dispersion than that derived observationally from assuming eMSTO features in IACs \citep{goud2014} are due to an age spread.  Such a discrepancy can be alleviated if assumed evolution of the velocity dispersion changes more dramatically from YMC to IAC stage than assumed in \cite{goud2014} or if the eMSTO feature is due to a population of rapidly rotating main sequence stars  \citep{cabrera2016,bastian2016,piatti2016} or other stellar evolutionary affects \citep{bastian2017}.
While \cite{li2016} see evidence for past star formation in IACs which are less massive and more diffuse clusters than predicted by our simulations, their suggestion that these episodes of star formation may have been triggered by the accumulation of gas from the clusters as they orbited within the gaseous disk of their host galaxy is not necessarily inconsistent with this work as we assume our stellar wind material is accumulated in star clusters in isolation.  Previous work has shown that cold gas accretion may indeed be a viable avenue for significant gas retention and star formation \citep{naiman2009,conroy2011a,conroy2011b,priestley2011,naiman2011,conroy2012}, however our detailed treatment of the interplay between these two gas accumulation processes is left to a subsequent paper. 

\subsubsection{Relationship to Evolved Stellar Clusters}

 Abundance variations of light elements in main sequence stars \citep{gratton2001,briley2002,cohen2002,cannon1998,pancino2010b} and RGB stars \citep{sneden2004} within the majority of Galactic globular clusters suggest a complex enrichment history during star formation.
 Our results pose problems for many of the current scenarios for explaining the complex star formation histories observed in many globular clusters under the assumption that currently observed YMCs will eventually evolve into systems like Galactic globular clusters.

 Of the several scenarios attempting to explain abundance variations in globular clusters, many rely on a first generation of stars polluting the interstellar gas with enriched material which is then incorporated into a subsequent generation of stars.
 One of the more popular scenarios which may meet these requirements is enrichment from a combination of intercluster gas from the winds of AGB stars and accreted unprocessed material \citep{prantzos2007,ventura2008a,ventura2008b,pflamm2009,dercole2010,conroy2011a}.  The proposed timeline  for this method proceeds from an initial episode of star formation into an initial clearing of gas from SNe \citep{dantona2004,dercole2008,conroy2011a}.  Winds from AGB stars are then effectively retained within the proto-GC, mix with an accumulation of pristine gas accreted from the cluster's surroundings, and form a subsequent population of stars.  As the possible mass lost from AGB stars has an upper limit of approximately 10\% of the initial mass of the cluster, a stellar mass loss of approximately 90\% of the initial population of stars is invoked in order to explain the roughly equal masses of first and generation stars observed in present day GCs \citep{dercole2008,conroy2011a,conroy2011b,conroy2012}.

 Typically it is assumed the optimum time for retention of anomalous material is $\sim 50 - 300$~Myrs during which material from stars in a mass range of $4 \lesssim M/M_\odot \lesssim 8$ necessary to reproduce the abundance ratios is injected into the cluster \citep{ventura2001,karakas2007,ventura2008a,ventura2008b} and be consistent with the $\lesssim$1~Gyr age spreads between the enhanced and unenhanced populations currently observed in globular clusters \citep{larsen2011,cabrera2014,cabrera2016,mart2018}.   
 Our results are in conflict with this scenario as we find only minimal ($\lesssim 2$\%) gas retention within the star clusters in the age range conducive to reproducing the abundance anomalies in the AGB scenario.  
 However, it is possible the optimum time range for pollution with material enriched by the necessary AGB stars may be modified given the differences between abundance ratios produced by different AGB models \citep{fenner2004,karakas2006,choi2008,bekki2007,ventura2008a,ventura2008b,doherty2014a,doherty2014b}.

\begin{figure*}
\centering\includegraphics[width=0.48\textwidth]{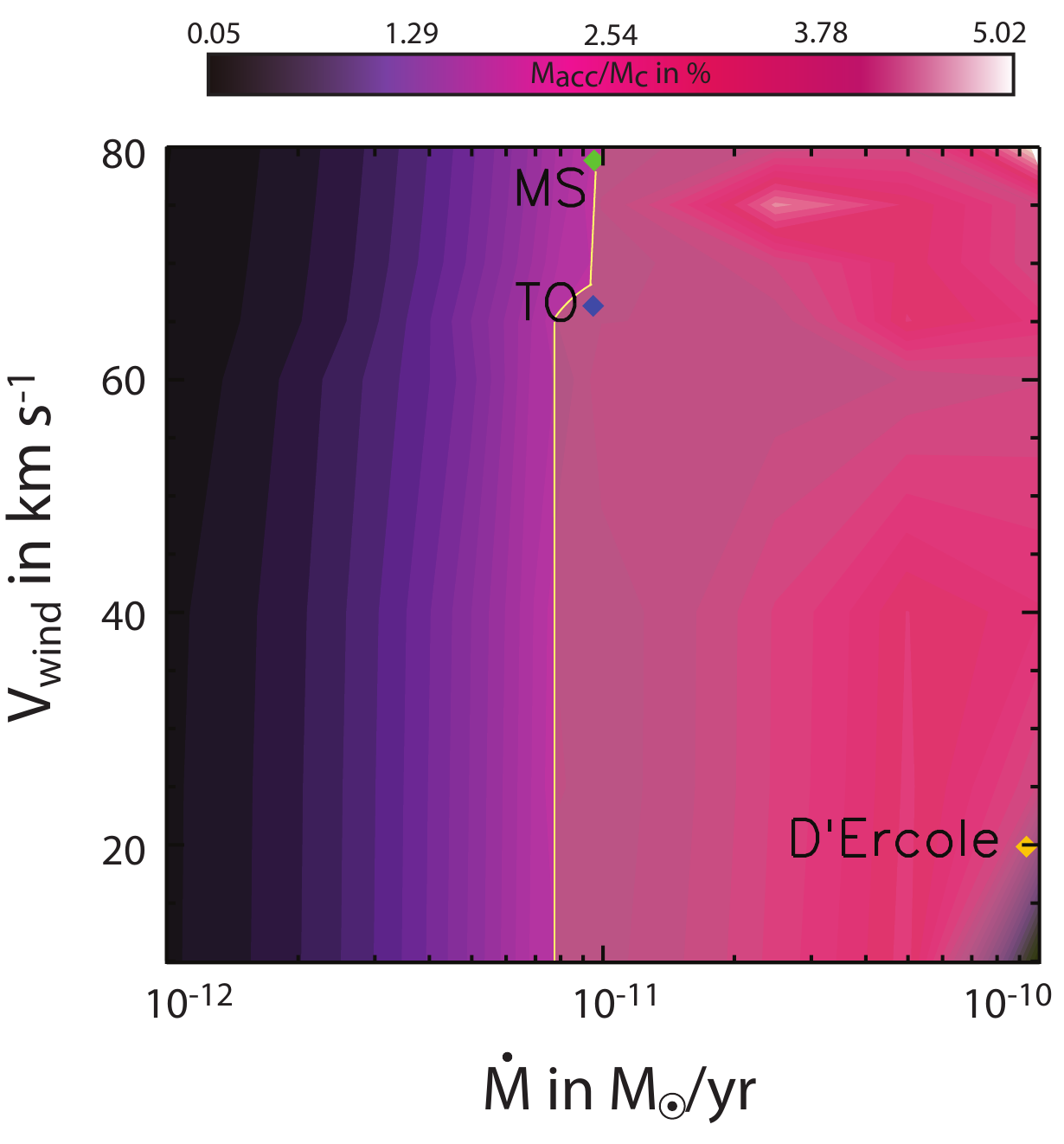}
\caption{Mass accumulated for a potential with $M_{\rm c} = 10^7 \, M_\odot$ and $\sigma_{\rm v} = 35 {\rm km s^{-1}}$ at 300~Myrs for different $\dot{M}$ and $v_{\rm w}$ combinations.  The shaded region to the right of the yellow line illustrates the parameter space in which star formation is triggered at 300~Myrs.  The colored diamonds denote the parameters which prescribe the mass loss rates and wind velocities from our turn off (blue) and population averaged prescriptions (green), along with the parameters from \citep{dercole2008} (yellow).}
\label{fig:mdotvw}
\end{figure*}

 Previous studies have shown greater gas mass retention is possible within YMCs in the proposed $\sim 50 - 300$~Myrs time span \citep{dercole2008,dercole2011,bekki2011}, however a smaller wind velocity and different star formation prescription was used in these calculations.
 We compare our mass loss and wind velocity prescription with the values derived from the work of \cite{dercole2008} in Figure \ref{fig:mdotvw}.  While the overall mass accumulation rate is smaller than in their work ($\sim 3-4$\% vs their $\sim 10$\%), we still find discrepancies in accumulation from the difference between assumed mass loss and wind velocity prescriptions.  The gas accumulation for the turn off prescription (labeled ``TO" in Figure \ref{fig:mdotvw}) is a factor of a few less than that in the D'Ercole $\dot{M}-v_w$ model, and the even lower rates observed from the population averaged prescription (``MS" in Figure \ref{fig:mdotvw}), the latter never cooling enough to trigger star formation in our models.

 Other scenarios proposed to explain the different abundances of subpopulations within present day GCs rely on complex star formation histories during a much smaller time range ($\sim 10-50$~Myrs) to explain the abundance anomalies observed in present day globular clusters
 \citep{maeder2006,prantzos2006,decressin2007a,decressin2007b,charbonnel2013,krause2013,bastian2013b,cassisi2014a,henault2015,milone2016,bastian2016}, with two of the most studied being the early disk accretion (EDA) and the fast rotating massive star and/or interacting binaries (FRMS/IBs) models.  
 The EDA scenario \citep[e.g.][]{bastian2013a} requires low mass stars to sweep up material pre-processed from rotating massive stars and/or interacting high mass binaries, while the FRMS/IBs scenarios \citep[e.g.][]{deMink2009,krause2013} assumes a second generation of stars forms from the ejecta of rapidly rotating and/or dynamically stripped massive stars within the first $20-40$~Myrs of a cluster's life.
 While only a small amount of material is expected to be retained by stellar winds during the time span in which these two scenarios operate ($0.5-2$\% between $\sim 10-50$~Myrs in Figures  \ref{fig:fig4} - \ref{fig:memcontour}), it is possible that it might have a mitigating effect on these processes.
 The EDA model requires the pre-main sequence disks to survive for $\sim 10-20$~Myrs.  Along with other disk-damaging effects \citep{scally2001,dejuan2012}, the existence of the hot, thermalized stellar wind intercluster medium could aid in the destruction of these pre-main sequence disks from ram pressure stripping during as the star moves through the YMC.
 Inclusion of hot stellar wind material during the first few tens of Myrs of a cluster's life also poses problems for the FRMS/IBs scenario.  If this small amount of stellar wind material efficiently thermalizes with the ejecta of FRMS or IBs, the additional heat might be able to prevent a second generation of star formation. Additionally, the combination of efficient mixing between stellar ejecta at early times could also lead to dispersions in Iron, which are not seen in all but a few clusters \citep[e.g.][]{villanova2007}.

 Other methods of generating a subpopulation of stars with enhanced light element abundances during the early phases of cluster evolution ($\lesssim$500~Myrs) which either rely on pollution sources that are highly centralized to the core of the proto-GC (Very Massive Stars, \citealt{den2014}; Extended Central Star Formation, \citealt{prantzos2006,elmegreen2017,wunsch2017}), a top heavy IMF \citep{charbonnel2014}, more complex gas dynamics during initial phases of star formation (Inhomogeneous Pre-enriched Gas, \citealt{marcolini2009}; Turbulent Elemental Separation, \citealt{hopkins2014}), or the reaccretion of remnant gas \citep{decressin2007a,decressin2007b} are not necessarily inconsistent with our models given that they depend on sources of enrichment that do not follow the stellar potential as we have assumed here.
 However, fast stellar winds would undoubtably add to the energy budget in such models, decreasing the gas retention potential of all scenarios.
 
 Indeed, the expected retention and expulsion of gas in YMCs as predicted from our stellar wind only models potentially poses problems for all of the popular explanations of globular cluster formation under the assumption that YMCs evolve into IACs and eventually globular clusters.
 Given our model's agreement with both the current observed gas content in YMCs (section \ref{section:YMCcurrent}) and observed ongoing star formation (section \ref{section:YMCongoing}), it is possible that the assumption that present day YMCs evolve into present day GCs needs to be relaxed.
 However, we caution that because current abundance models for all scenarios fail to fully reproduce the distribution of observed abundance anomalies \citep{bastian2015a}, the idea of a single formation pathway and any effects our stellar wind models may have on its success will likely need to be reevaluated as these scenarios evolve.

\subsection{Prospects for the Detection of Gas and Star Formation in Intermediate Age Clusters: Age $\sim$1-4~Gyrs} \label{section:IACs}

 Our models predict the existence of large gas reservoirs during the time period of 1-4~Gyrs in all clusters with $\sigma_{\rm v} \gtrsim 25 \, {\rm km s^{-1}}$ and $M \gtrsim 10^6 \, M_\odot$.
 In addition, if the main sequence winds thermalize inefficiently with the winds from the turn off stars (Figure \ref{fig:fig4}) or the gas retention in the cluster ``has memory" (Figure \ref{fig:memcontour}), a small amount of star formation is predicted to be triggered during this time span - approximately $\lesssim 10$\% of the cluster mass may be available for a second generation of star formation.  
 Assuming the generous rate of 10\% star formation efficiency, this implies only $\sim 1$\% of the cluster's mass forms stars once the cluster is older than $\sim$1-3~Gyrs.
 To observe such a small rate of star formation at high precision, we are limited in observations to within the Local Group ($\sim$1-2~Mpc) as photometrically derived SFHs in general to agree with CMD derived ones at a level of a few percent, a level which is far higher then the $\lesssim 1$\% one would need to resolve \citep{ruiz2015}.
 In one of the largest searches for star formation locally in the Magellanic Clouds, no ongoing star formation has been observed in star clusters with ages up to $\sim$1~Gyr \citep{bastian2013a}.
 For observations of older clusters matters are further complicated as to accurately age data clusters photometrically in the range of 1-4~Gyrs both optical and NIR are necessary to break the strong age-metallicity degeneracy between these and the much older $\sim$13~Gyr old clusters \citep{trancho2014}, making such observations more expensive for clusters older than $\sim$1~Gyr.

 Merging galaxies provide another possible avenue to observe ongoing star formation and gas retention in massive clusters, albeit at lower precision than in the Local Group, as these systems typically generate a multitude of star cluster formation with ages approximately the age of the merger itself \citep{whitmore2007}.
 However observations of one of the most studied mergers, the Antennae galaxies, occurring $\sim$500 Myrs ago, shows that while upper limits on gas content show intercluster gas at the level of $\lesssim$9\% \citep{cabrera2015}, there are large gas reservoirs surrounding the clusters \citep{zhu2003} - hardly the clusters in isolation studied here - making a strict comparison between our prescription and those in intermediate age merging systems less straight forward.  
 Furthermore, external gas reservoirs could either aid in the retention of intercluster gas or help with its expulsion.  If the clusters move through hot and/or dense gas ram pressure stripping may be effective at removing material over a few sound crossing times across the cluster \citep{frank1976,naiman2009,desilva2009,martell2009,pancino2010a,naiman2011,priestley2011}.  
 If, however, they move slowly enough relative to the sound speed of the gas the large reservoir of external material may aid in the retention of gas within the clusters' centers \citep{pflamm2009,naiman2011,priestley2011}.

 More locally, the effects of gas retention on observations of SMC and LMC star clusters has focused predominately on the YMC population \citep{bastian2014a,longmore2015}.  
  While there are IACs in the LMC and SMC with large extinctions there is not an overall trend with age \citep{perren2017} as would be suggested by the work presented here.
 However, we caution here that once again LMC and SMC clusters are not necessarily the isolated systems discussed in this work.

 Finally, if IACs are believed to evolve into GCs, the amounts of gas retained from stars during the 1-4~Gyr time span would possibly lead to subsequent populations of stars with enhanced C+N+O, which is not currently observed in the enhanced population of stars in galactic GCs \citep{decressin2009}.

 While observations of star formation and gas retention in IACs are far from complete, the observed lack of star formation and gas retention implies that there is efficient thermalization between main sequence and evolved stellar winds as depicted in Figure \ref{fig:fig5} or an additional source of heat in these systems.
 Additionally, other mechanisms beyond ram pressure stripping and stellar wind feedback may be responsible for clearing the gas.  
 In lower mass and less dense clusters, radiation from white dwarfs may result in the ionization of the gas at ages $\gtrsim$1~Gyr \citep{mcdonald2015}, and classical novae may be able to drive gas out in the lower mass clusters \citep{moore2011}.  
 Furthermore, in both low and high mass clusters, the accretion onto stellar-mass black holes could aid in removing this build up of mass in the system \citep{leigh2013a}.  

 Both further observations of star formation and gas content in IACs and more physically realistic simulations are needed before a more robust comparison between the stellar wind retention models presented here and the observed properties of IACs can be made.

\subsection{Current Gas Content in Globular Clusters: Age $\gtrsim 7$~Gyrs}

 As there is little to no ongoing star formation predicted in our models and observed in galactic GCs or old SMC and LMC clusters \citep[e.g.][]{mart2018}, we here focus solely on the current gas content of globular clusters.

 Given that the typical time between gas stripping Galactic GCs' disk passages, 0.1~Gyrs \citep{odenkirchen1997}, is long enough for significant slow RGB wind material to be injected into GCs \citep{tayler1975}, if this wind material is effectively retained within the cluster we expect to see approximately 10-100~$M_\odot$ of gas within Galactic GCs.
 However, the majority of Galactic GCs show little or no evidence for gas of 10-100~$M_\odot$ within their interiors, as measured either in neutral hydrogen \citep{heiles1966, robinson1967, kerr1972, knapp1973b, bowers1979, birkinshaw1983, lynch1989, smith1990, vanloon2006, vanloon2009}, CO \citep{troland1978,smith1995, leon1996}, OH and $\rm{H_2 O}$ maser emission \citep{knapp1973a, kerr1976, frail1994, cohen1979, dickey1980, vanloon2006}, or dust \citep{lynch1990, knapp1995, origlia1996, hopwood1998,hopwood1999,evans2003, boyer2006,barmby2009}.  
 The most stringent constraints come from a combination of radio dispersion measurements of millisecond pulsars in 47~Tucanae \citep{camilo2000} showing $n_{\rm e} = 0.067 \pm 0.015 \, {\rm cm^{-3}}$ and neutral hydrogen measurements indicating $M_{\rm HI} \lesssim 3.7 \, M_\odot$, which when taken together indicate a derth of gas within 47~Tucanae.
 In addition, as mentioned in previously in section \ref{section:IACs}, while there are enhancements in extinction in SMC and LMC clusters, it is not a highly ubiquitous feature of the old clusters in these systems \citep{perren2017}.

 In Figures \ref{fig:fig4} - \ref{fig:memcontour} we find little expected gas retention in all but the most compact star clusters ($\sigma_{\rm v} \lesssim 40-45 \, {\rm km s^{-1}}$) at a population age of 10~Gyrs, but caution any gas retention is likely to decrease for less massive clusters ($M_{\rm c} < 10^7 \, M_\odot$) as indicated in the right hand panel of Figure \ref{fig:fig4}.  For $\sigma_{\rm v} \lesssim 23 \, {\rm km s^{-1}}$ we find negligible mass retention across all mass loss prescriptions and cluster masses ($M_{\rm acc}/M_{\rm c} \sim 0\%$).  
 Thus, our results are in excellent agreement with both observations and a previous analytically derived limit of $\sigma_{\rm v} \lesssim 22 \, {\rm km s^{-1}}$ by \cite{smith1999} who also account for the effects of main sequence stellar wind heating on the intercluster gas.

\section{Summary and Conclusions} \label{section:conclusions}

 We have presented a grid of spherically symmetric simulations of gas retention and expulsion due to stellar winds within star clusters.
 While previous work has estimated the ability of star clusters to retain stellar winds \citep{dercole2008,dercole2010,vesperini2010,conroy2011b,conroy2011a}, calculations have so far been under taken over a smaller range of cluster properties and stellar ages, typically with lower stellar wind velocities than those argued for in this work.
 Motivated by this, we have calculated gas retention in  star clusters of various ages, stellar mass and compactness.  
 Additionally, we include a discussion of the choice of the kinetic energy injection proxy, taken here as the wind velocity, $v_w$, and its relation to the observed distribution of outflow velocities observed in field and globular cluster AGB and RGB stars.  \\
 \\
\noindent Our main conclusions are the following:

\begin{easylist}[itemize]

& We conclude that in compact star clusters a choice of $v_w = v_{\rm esc}$ best reproduces both the observations of stellar winds and the assumed level of mixing between stellar wind and the intercluster medium.

& Given our assumptions about the mass loss rates and wind velocities emanating from stars residing in clusters, we find the optimum time for gas retention is approximately 1-2~Gyrs.  However, depending on how efficiently the slow winds from AGB/RGB stars mix with the fast winds from the numerous main sequence cluster members, and how effectively gas is cleared after each episode of star formation within the cluster, we find star formation may not be triggered within this optimum time span as the gas may be too hot.

& We find significant gas retention can occur when cluster velocity dispersions are high, $\sigma_{\rm v} \gtrsim 25 \, {\rm km s^{-1}}$, but the amount of gas retained drops as the velocity dispersion grows due to the efficient funneling of gas to the central regions of the cluster and subsequent triggering of star formation for very compact clusters.

& We compare our results with observations of young massive clusters (YMCs), intermediate age clusters (IACs) and globular clusters (GCs).  We find our models generally agree with observations.  We predict little gas retention and star formation in YMCs and GCs.  Our models predict higher levels of gas retention in IACs ($\approx 10\%$ of the stellar mass).  However, we caution that both the lack of observations of isolated IAC systems and the lack of inclusion of all the sources of thermal and kinetic energy in star clusters in this age range makes any conclusions drawn from our models preliminary.  

& Finally, we discuss the implications of our models on the assumed evolutionary sequence of YMC$\rightarrow$IAC$\rightarrow$GC resulting in observations of multiple sub-populations with different levels of light element enhancement in present-day GCs.  The majority of proposed origins for these sub-populations rely on some amount of gas retention in the age range few-100s of Myrs. However, we find hot stellar winds are effective at driving material from the cluster during this time frame for all but the most massive and compact clusters.  This result is problematic for the AGB wind retention scenario \citep{dercole2008,dercole2010,conroy2011a,conroy2011b} invoked to explain the abundance anomalies observed in present day Galactic globular clusters, but if hot stellar winds can effectively thermalize with the intercluster medium during this time, our models pose problems for gas retention in the early disk accretion \citep{bastian2013a} and fast rotating massive stars \citep{krause2013} scenarios as well.

\end{easylist}

$\,$\\

\noindent In this work, we have also tried to minimize the effect of external mass inflow by considering star clusters in isolation. This is certainly not the case for clusters moving through cold, dense environments, as they  can potentially 
amass a significant amount of gas from their surroundings \citep{naiman2009,naiman2011}, or clusters moving quickly though hot halo gas or the galactic disk \citep{priestley2011}.  If the star clusters reside  within cold gas, stellar winds and exterior inflows in such clusters could combine 
to create even larger central density enhancements \citep{naiman2009,pflamm2009,naiman2011,conroy2011a}.  
Additionally, one dimensional spherically symmetric models cannot accurately model all multi-dimensional phenomena. For example, these simulations are not able to follow the effects of cool fragments which may exist in otherwise hot gas, thereby overestimating the effects of hot gas expelling material from the cluster \citep{krause2012}.
A  self consistent  treatment  of both interior and exterior gas  accumulation in multi-dimensional simulations,  which includes the effects of compact object accretion, tidal stripping, photoionization, and pulsar heating will be presented elsewhere.


\section*{Acknowledgements}

We thank Mark Krumholz, Charlie Conroy, David Pooley, Rebecca Bernstein, Chung-Pei Ma and Morgan Macleod for useful discussions, and we thank the referee for their helpful comments. The software used in this work was in part developed by the DOE-supported ASCI/Alliance Center for Astrophysical Thermonuclear Flashes at the University of Chicago. Computations were performed on the Pleaides/Hyades UCSC computer cluster. This work is supported by NSF: AST-0847563, NASA: NNX08AL41G, NSF AARF award AST-1402480, and The David and Lucile Packard Foundation.

\bsp	





\bibliographystyle{mnras}

\bibliography{bib_agb_2016}


\label{lastpage}
\end{document}